\documentclass[manuscript,screen,review=false]{acmart}

\usepackage{flushend}
\usepackage{balance}
\usepackage{multirow}
\usepackage{booktabs}
\usepackage{graphicx}

\usepackage{url}       
\usepackage{stfloats}  
\usepackage{amsmath}   
\usepackage{array}
\usepackage{breqn}
\usepackage{amsfonts}
\usepackage{courier}

\usepackage{amssymb}
\usepackage{balance}
\usepackage{multirow,tabularx}
\usepackage{longtable}
\usepackage{booktabs,longtable}
\usepackage{hhline}
\usepackage{dsfont}
\usepackage{fancyhdr}
\usepackage{epstopdf}
\usepackage{algorithm}
\usepackage{enumitem}
\usepackage{algorithmic}
\usepackage[bottom]{footmisc}
\usepackage{multicol}
\usepackage{multirow}
\usepackage{verbatim}
\usepackage{comment}
\usepackage{longtable}
\usepackage{arydshln}
\usepackage{color} 
\usepackage{xcolor}
\usepackage{caption}
\usepackage{makecell}

\usepackage{tikz}
\usepackage{forest}
\usepackage{mydef}

\usepackage[most]{tcolorbox}

\useforestlibrary{edges}

\AtBeginDocument{%
  }

\setcopyright{acmlicensed}
\copyrightyear{2024}
\acmYear{2024}
\acmJournal{TOIS}
\acmDOI{XXXXXXX.XXXXXXX}

\acmISBN{978-1-4503-XXXX-X/18/06}

\begin{document}

\title{Graph Foundation Models for Recommendation: A Comprehensive Survey}

\author{Bin Wu}
\orcid{0009-0004-2162-4610}
\affiliation{%
    \institution{Beijing University of Posts and Telecommunications}
    \city{Beijing}
   \country{China}
}
\email{wubin0354@gmail.com}

\author{Yihang Wang}
\orcid{0009-0006-3819-2193}
\affiliation{%
    \institution{University of Chinese Academy of Sciences}
    \city{Beijing}
   \country{China}
}
\email{yihangwang1020@gmail.com}

\author{Yuanhao Zeng}
\orcid{0009-0009-6213-573X}
\affiliation{%
    \institution{ShanghaiTech University}
    \city{Shanghai}
   \country{China}
}
\email{lineshogan@gmail.com}

\author{Jiawei Liu}
\orcid{0000-0003-2437-0455}
\affiliation{%
    \institution{Beijing University of Posts and Telecommunications}
    \city{Beijing}
   \country{China}
}
\email{liu_jiawei@bupt.edu.cn}

\author{Jiashu Zhao}
\orcid{0000-0001-9770-7616}
\affiliation{%
    \institution{Wilfrid Laurier University}
    \city{Waterloo}
   \country{Canada}
}
\email{jzhao@wlu.ca}

\author{Cheng Yang}
\orcid{0000-0001-7821-0030}
\affiliation{%
    \institution{Beijing University of Posts and Telecommunications}
    \city{Beijing}
   \country{China}
}
\email{yangcheng@bupt.edu.cn}

\author{Yawen Li}
\orcid{0000-0003-2662-3444}
\affiliation{%
    \institution{Beijing University of Posts and Telecommunications}
    \city{Beijing}
   \country{China}
}
\email{warmly0716@126.com}

\author{Long Xia}
\orcid{0009-0007-7536-6241}
\affiliation{%
    \institution{Baidu Inc.}
    \city{Beijing}
   \country{China}
}
\email{long.phil.xia@gmail.com}

\author{Dawei Yin}
\orcid{0000-0002-0684-6205}
\affiliation{%
    \institution{Baidu Inc.}
    \city{Beijing}
   \country{China}
}
\email{yindawei@acm.org}

\author{Chuan Shi}
\authornote{Corresponding author.}
\orcid{0000-0002-3734-0266}
\affiliation{%
    \institution{Beijing University of Posts and Telecommunications}
    \city{Beijing}
   \country{China}
}
\email{shichuan@bupt.edu.cn}

\renewcommand{\shortauthors}{Bin Wu et al.}
\thanks{This work has been submitted to the ACM for possible publication. Copyright may be transferred without notice, after which this version may no longer be accessible.}

\begin{abstract}
    Recommender systems (RS) serve as a fundamental tool for navigating the vast expanse of online information, with deep learning advancements playing an increasingly important role in improving ranking accuracy. Among these, graph neural networks (GNNs) excel at extracting higher-order structural information, while large language models (LLMs) are designed to process and comprehend natural language, making both approaches highly effective and widely adopted. Recent research has focused on graph foundation models (GFMs), which integrate the strengths of GNNs and LLMs to model complex RS problems more efficiently by leveraging the graph-based structure of user-item relationships alongside textual understanding. In this survey, we provide a comprehensive overview of GFM-based RS technologies by introducing a clear taxonomy of current approaches, diving into methodological details, and highlighting key challenges and future directions. By synthesizing recent advancements, we aim to offer valuable insights into the evolving landscape of GFM-based recommender systems.
\end{abstract}




\maketitle

\section{Introduction}
Recommender systems (RS) have become indispensable in modern online platforms, helping users navigate information overload by proactively surfacing personalized content, products, and services in scenarios such as e-commerce, social media, and entertainment~\cite{adomavicius2005toward,isinkaye2015recommendation,su2009survey}.
At their core, RS aim to model user preferences (or utilities) over a large candidate space and learn a ranking policy that matches users with items they are most likely to engage with under contextual constraints~\cite{rendle2012bpr,covington2016deep}.
The efficacy of such recommendation policies fundamentally depends on learning accurate representations from the data sources available in the recommendation environment~\cite{su2009survey,isinkaye2015recommendation}.
Consequently, in practice, RS are typically tasked with leveraging heterogeneous and often sparse signals that intertwine \emph{structural information} (e.g., user-item interactions and relational dependencies) with \emph{textual information} (e.g., user profiles, item descriptions, and reviews)~\cite{van2000using,su2009survey,liu2023understanding}.
Effectively integrating these sources is central to improving relevance, robustness, and generalization in real-world recommendation.

\begin{figure}[ht]
    \centering
    \includegraphics[width=0.8\linewidth]{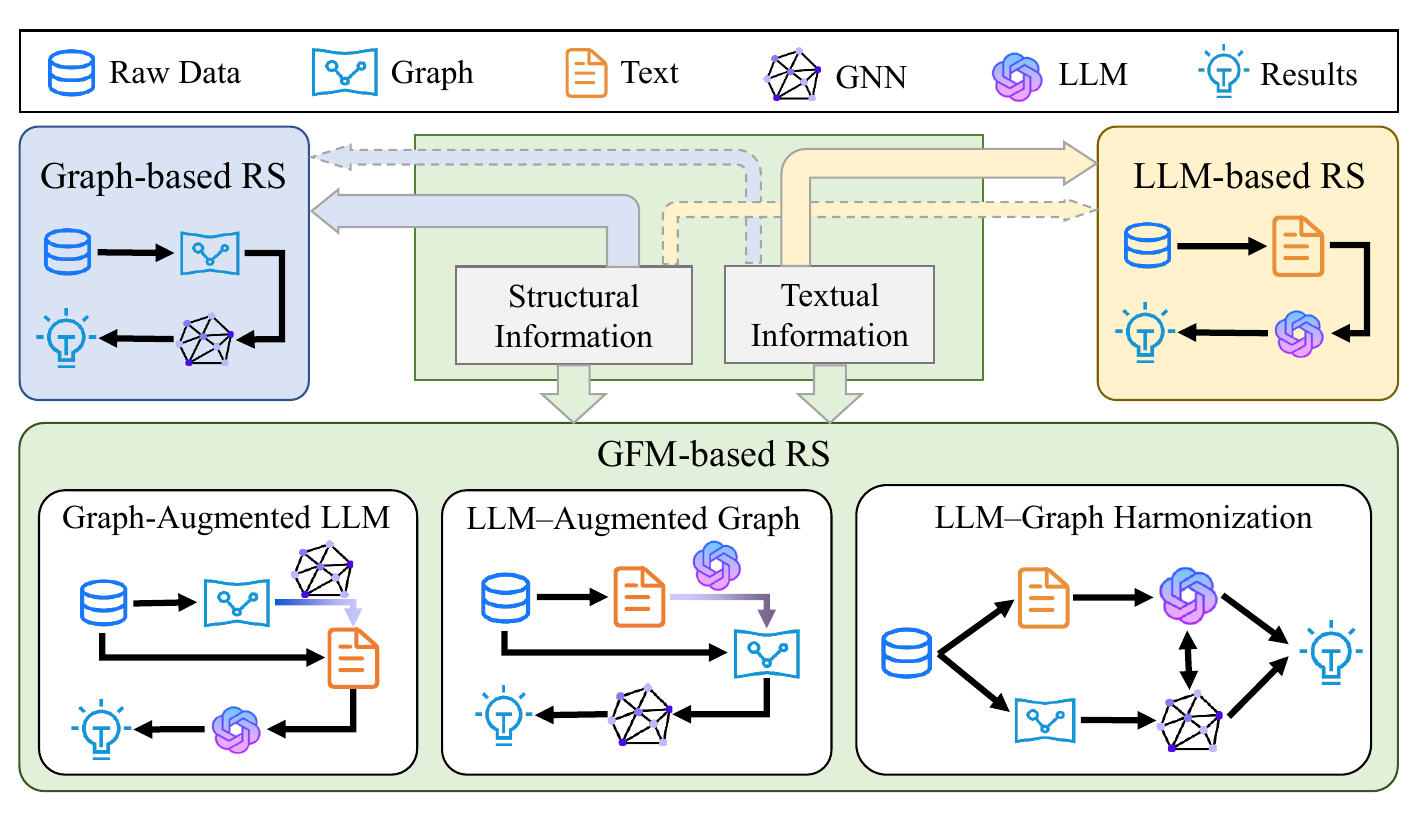}
    \caption{An overview of GFM-based RS. Compared with GNN-based or LLM-based RS, GFM-based RS are positioned as integrating both approaches to create more comprehensive recommendations.}
    \label{fig:overview}
\end{figure}

These characteristics continue to motivate RS modeling, and deep learning has been a major driver of progress, particularly in how models represent and exploit structural and textual evidence.
As illustrated in Figure~\ref{fig:overview}, RS data primarily consists of structural and textual information, yet graph neural networks (GNNs) and large language models (LLMs) exhibit varying efficiencies in utilizing these two modalities.
On the one hand, GNN-based RS have become a dominant paradigm for leveraging collaborative information from interaction data, where message passing enables the aggregation of higher-order dependencies and can incorporate node attributes, but may be limited by how unstructured language is compressed and aligned with message passing~\cite{wu2022graph,he2020lightgcn,wang2019neural}.
Yet, purely graph-centric inductive biases make it non-trivial to fully exploit rich natural language associated with users and items.
On the other hand, LLM-based RS excel at processing text through sequential modeling, leveraging contextual representation learning and world knowledge to understand and reason over natural language~\cite{yang2023palr,zhai2024actions,wu2024survey,lin2025can}, yet this paradigm can overlook or dilute collaborative filtering signals that are not naturally expressed in token sequences.
However, due to the lack of an explicit mechanism for relational propagation, LLMs alone often struggle to model higher-order structural dependencies and consistently exploit interaction graphs for recommendation~\cite{wei2024llmrec,guo2024integrating}.
With carefully designed frameworks, GNNs and LLMs can exchange task-relevant sparse signals that are often implicit and difficult to obtain in isolation across their respective views.
In particular, structural cues distilled by GNNs can be injected into LLMs in token-compatible forms, while semantic priors distilled by LLMs can be aligned to graph representations and exploited by message passing.
This motivates graph foundation models (GFMs), which synergistically couple GNNs and LLMs within a unified framework to jointly model graph structures and language semantics~\cite{liu2023towards,mao2024position,mao2024graph,ren2024survey,jin2024large}.
From a data perspective, such a unification matches the needs of RS to leverage more heterogeneous data and improve data utilization, jointly modeling structural signals and textual semantics that are often underexploited when treated in isolation.
GFM-based RS offer a promising paradigm for aligning user preferences with both relational evidence and textual semantics, potentially yielding more accurate and robust recommendations.

To better understand this emerging paradigm and its design space, we summarize GFM-based RS through a taxonomy that captures how graphs and LLMs are coupled in practice.
To systematize this rapidly growing literature, we propose a taxonomy based on the \emph{synergistic relationship} between graphs and LLMs in GFMs~\cite{ren2024survey}: \textbf{Graph-augmented LLM}, \textbf{LLM-augmented Graph}, and \textbf{LLM-Graph harmonization}.
\textbf{Graph-augmented LLM} injects graph-derived structural cues into LLMs to enhance recommendation reasoning and generation, as exemplified by LLMGR~\cite{guo2024integrating}, which feeds GNN-learned embeddings into the token sequence and adapts the model via two-stage fine-tuning.
\textbf{LLM-augmented Graph} uses LLMs to enrich or reshape graph information (e.g., by constructing auxiliary structures or extracting semantic signals) before graph-based encoding, as exemplified by LLMRG~\cite{wang2023enhancing}, which builds inference and divergence graphs from interaction histories and then applies GNNs for recommendation.
\textbf{LLM-Graph harmonization} aims to align and jointly optimize structural and semantic representations in a shared space, as exemplified by DALR~\cite{peng2024denoising}, which aligns GNN-encoded and LLM-encoded embeddings for downstream recommendation.

As an evergreen topic in both academia and industry, RS have been the subject of numerous surveys (e.g., \cite{gao2023survey,wu2024survey,liu2023towards,li2023survey}). \cite{gao2023survey,wu2024survey} focus on specific methodologies, such as GNN-based RS or the more recent LLM-based RS. \cite{li2023survey} concentrates on utilizing LLM to enhance graphs for tackling tasks related to graphs. However, the field is rapidly evolving with GFMs emerging as a crucial technique of the RS research. \cite{liu2023towards} systematically outlines the existing GFMs from the perspectives of pre-training and adaptation, while overlooking the recommendation which is one of the significant downstream tasks for GFM. This survey provides a timely and comprehensive overview that covers the landscape of GFM-based recommender systems.

The contributions of this survey can be summarized in the following aspects:
\begin{itemize}[topsep=0pt]
    \item \textbf{Pioneering overview}: Our survey fills the blank in comprehensive work in the field of GFM-based RS.
    \item \textbf{Clear taxonomy}: The comprehensive survey presents a well-structured taxonomy of GFM-based RS, allowing future work to be easily categorized within the corresponding branches.
    \item \textbf{Promising outlook}: We present the challenges and future research directions in this field, which can serve as a valuable reference for research in this rapidly evolving area.
\end{itemize}

\section{Background and Preliminaries}
This section reviews the recommendation setting from a representation and supervision perspective, and then clarifies what we mean by graph foundation models (GFMs) in the context of recommender systems. Building on these foundations, we explain why coupling graph-structured collaborative evidence with natural-language semantics is a recurring requirement across the evolution of recommender systems, and how GFM-style designs respond to this requirement in practice. We finally summarize the design space with a taxonomy based on the primary coupling direction between graphs and LLMs.

\subsection{Recommender Systems}

Recommender systems (RS) are designed to match each user with a small set of relevant items from an extremely large candidate space~\cite{su2009survey,adomavicius2005toward}, based on observed signals such as clicks, purchases, ratings, and other forms of implicit feedback~\cite{hu2008collaborative,rendle2012bpr}. In practice, the supervision of recommendation is interaction-driven: models learn from user--item behaviors that are sparse, noisy, heavily long-tailed and biased due to exposure effects~\cite{schnabel2016recommendations,joachims2017unbiased}, while online serving requires efficient retrieval and ranking under strict latency and memory constraints~\cite{covington2016deep,huang2025comprehensive}. This combination makes representation learning central to RS: the core question is not only how to fit observed interactions, but also how to generalize from limited feedback to unseen items, cold-start entities, and evolving user interests.

From a data abstraction perspective, most recommendation environments naturally induce graph-structured evidence. User-item interactions can be viewed as a bipartite graph~\cite{baluja2008video}, and this interaction graph is often extended into heterogeneous graphs by incorporating side information such as categories, brands, creators, tags, or knowledge relations~\cite{wang2018ripplenet,wang2019kgat,shi2020heterogeneous}. Meanwhile, users and items are increasingly accompanied by rich natural-language fields (profiles, titles, descriptions, reviews, comments, and conversations), which serve as high-capacity carriers of semantics and prior knowledge but are not directly aligned with interaction supervision~\cite{liu2023understanding,li2018towards,hayati2020inspired}. As a result, modern RS data can be described as a heterogeneous graph in which nodes and edges encode relational structure, while textual content appears as node/edge attributes that must be transformed into usable representations.

Historically, RS have progressed from methods that primarily rely on IDs and co-occurrence statistics toward methods that attempt to leverage broader evidence. Classical collaborative filtering and matrix factorization focus on learning latent factors from interaction matrices~\cite{koren2009matrix,su2009survey}, while content-based methods incorporate item attributes to alleviate sparsity. With the rise of deep learning, recommender models expanded to capture non-linear feature interactions~\cite{he2017neural,guo2017deepfm} and sequential dynamics~\cite{sun2019bert4rec,hidasi2015session,kang2018self}, improving accuracy but still facing a recurring bottleneck: interaction signals alone are often insufficient to explain user preferences when feedback is sparse, and side information must be utilized in a way that remains faithful to interaction-driven relevance~\cite{covington2016deep,schnabel2016recommendations}. 

This bottleneck reflects a persistent mismatch between \emph{how recommendation supervision is formed} and \emph{how rich semantic evidence is expressed}. Collaborative signals emerge from collective behaviors and higher-order dependencies in interaction graphs; they are powerful but implicit and sparse, and often require relational aggregation to become predictive~\cite{wang2019neural,he2020lightgcn}. Meanwhile, natural-language content provides expressive semantics and transferable prior knowledge, yet semantic relatedness does not automatically translate into preference relevance under long-tail exposure and feedback biases~\cite{schnabel2016recommendations,joachims2017unbiased}. Therefore, modern RS increasingly require representation learning mechanisms that can jointly exploit relational structure and semantic evidence while keeping interaction-driven relevance as the optimization anchor.

\subsection{Graph Foundation Models}

Building on the above mismatch between interaction-driven supervision and rich semantic evidence, we define \emph{graph foundation models (GFMs)}~\cite{mao2024position,wang2025graph} in recommender systems as an RS-oriented paradigm that jointly leverages (i) graph-structured collaborative signals and (ii) foundation-model semantic capability~\cite{bommasani2021opportunities}. In this RS-oriented view, the interaction graph provides an explicit substrate to aggregate higher-order dependencies and relational constraints, while large language models (LLMs) contribute transferable semantic understanding and prior knowledge from textual side information (e.g., titles, descriptions, reviews, and profiles). The key objective is to integrate these complementary sources in a way that remains anchored to user--item interactions, since recommendation relevance is ultimately defined by behavior-driven supervision rather than semantic similarity alone; meanwhile, the resulting representations and inference procedures must still be compatible with large-scale retrieval/ranking under strict efficiency constraints.

We further distinguish \emph{general-purpose GFMs} from \emph{GFM-based recommender systems} and adopt the latter in a deliberately broad sense throughout this survey. In the general-purpose interpretation, GFMs follow the foundation-model paradigm by pre-training on large-scale graph corpora and then adapting to diverse downstream tasks, aiming to learn reusable representations that jointly capture structural regularities (e.g., connectivity and multi-hop dependencies) and attribute semantics (e.g., textual node/edge descriptions)~\cite{liu2023towards,wang2025graph}. However, many representative recommendation methods do not strictly instantiate such a graph-pretrained backbone. Instead, they typically build upon pre-trained foundation components (most commonly LLMs) and design mechanisms to preserve, inject, or align interaction-driven collaborative evidence with semantic and structural representations under recommendation objectives. Accordingly, in this survey, we use "GFM-based recommender systems" to cover RS methods that explicitly leverage \emph{pre-trained model capability} and \emph{integrate both LLM- and GNN-style modeling}, regardless of whether the graph component is broadly pre-trained or mainly trained/adapted within the recommendation domain.

\subsection{GFM-based RS}

Under the above RS-oriented definition, GFM-based methods are best understood as a design space rather than a single uniform model family~\cite{huang2025survey,wu2024survey,lin2025can}. Their shared objective is to make pre-trained foundation capability practically usable under recommendation supervision and constraints: preference learning is anchored by user--item interactions, yet user and item understanding increasingly depends on natural language and other side information; meanwhile, deployment requires efficiency over large candidate spaces. Consequently, existing work differentiates itself mainly by deciding where the foundation components sit in the pipeline and how structural information and textual information are coupled during training and inference, which naturally leads to the taxonomy introduced next.

\tikzstyle{leaf}=[draw=hiddendraw,
    rounded corners,minimum height=1em,
    fill=mygreen!40,text opacity=1, align=center,
    fill opacity=.5,  text=black,align=left,font=\scriptsize,
    inner xsep=3pt,
    inner ysep=1pt,
    ]
\tikzstyle{middle}=[draw=hiddendraw,
    rounded corners,minimum height=1em,
    fill=output-white!40,text opacity=1, align=center,
    fill opacity=.5,  text=black,align=left,font=\scriptsize,
    inner xsep=3pt,
    inner ysep=1pt,
    ]
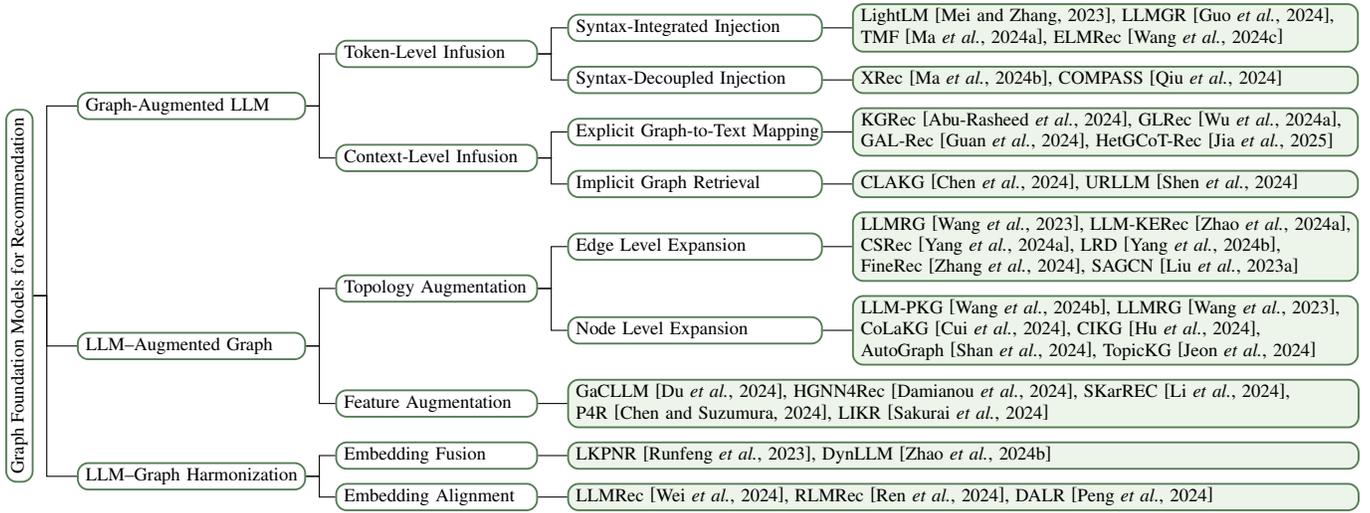
\begin{figure*}[ht]
\centering
\begin{forest}
  for tree={
  forked edges,
  grow=east,
  reversed=true,
  anchor=base west,
  parent anchor=east,
  child anchor=west,
  base=middle,
  font=\scriptsize,
  rectangle,
  line width=0.7pt,
  draw=output-black,
  rounded corners,align=left,
  minimum width=2em,
    s sep=5pt,
    inner xsep=3pt,
    inner ysep=1pt,
  },
  where level=1{text width=4.5em}{},
  where level=2{text width=6em,font=\scriptsize}{},
  where level=3{font=\scriptsize}{},
  where level=4{font=\scriptsize}{},
  where level=5{font=\scriptsize}{},
  [Graph Foundation Models for Recommendation, middle,rotate=90,anchor=north,edge=output-black
    [Graph-Augmented LLM, middle, edge=output-black,text width=8em
        [Token-Level Infusion, middle, text width=7em, edge=output-black
            [Syntax-Integrated Injection, middle, text width=9em, edge=output-black
                [LightLM \cite{mei2023lightlm}{,} LLMGR \cite{guo2024integrating}{,} TMF \cite{ma2024triple}{,} ELMRec \cite{wang2024enhancing}, leaf, text width=14.5em, edge=output-black]
            ]
            [Syntax-Decoupled Injection, middle, text width=9em, edge=output-black
                [XRec \cite{ma2024xrec}{,} COMPASS \cite{qiu2024unveiling}{,} Laser \cite{zhang-etal-2025-bi}, leaf, text width=14.5em, edge=output-black]
            ]
        ]
        [Context-Level Infusion, middle, text width=7em, edge=output-black
            [Explicit Graph-to-Text Mapping, middle, text width=9em, edge=output-black
                [KGRec \cite{abu2024knowledge}{,} GLRec \cite{wu2024exploring}{,} GAL-Rec \cite{guan2024enhancing}{,} CGR \cite{yang2025personalized}{,} \\HetGCoT-Rec \cite{jia2025hetgcot}{,} K-RagRec \cite{wang-etal-2025-knowledge-graph}{,} CausalX \cite{li2025counterfactual}{,} \\HeLLM \cite{pv3b-bx85-25}{,} MSCRS \cite{wei2025mscrs}{,} KERL \cite{mohbat2025kerl}{,} R2Rec \cite{zhao2025reason}{,} \\CORONA \cite{chen2025corona}{,} LLM-ER \cite{xu2025llm}, leaf, text width=14.5em, edge=output-black]
            ]
            [Implicit Graph Retrieval, middle, text width=9em, edge=output-black
                [RETURN\cite{ning2025retrieval}{,} ColdRAG \cite{yang2025cold}{,} OptiRAG-Rec \cite{zhou-etal-2025-efficiency}{,} \\G-Refer \cite{li2025g}{,} G-CRS \cite{qiu2025graph}{,} CLAKG \cite{chen2024leverage}{,} URLLM \cite{shen2024exploring}{,} \\KERAG_R \cite{meng2025kerag_r}{,} VoteGCL \cite{nguyen2025votegcl}, leaf, text width=14.5em, edge=output-black]
            ]
        ]
    ]
    [LLM–Augmented Graph, middle, edge=output-black, text width=8em
        [Topology Augmentation, middle, text width=7em, edge=output-black
            [Edge Level Expansion, middle, text width=9em, edge=output-black
                [HGPO \cite{luo2025llm}{,} LLM-KERec \cite{zhao2024breaking}{,} CSRec \cite{yang2024common}{,} LRD \cite{yang2024sequential}{,} \\FineRec \cite{zhang2024finerec}{,} SAGCN \cite{liu2023understanding}{,} Edu-Rec \cite{abu2025llm}{,} TagCF \cite{yuyou}{,} \\CKG-LLMA \cite{cai2025boosting}{,} LLMRG \cite{wang2023enhancing}{,} TWISTER \cite{wang2025towards}{,}, leaf, text width=14.5em, edge=output-black]
            ]
            [Node Level Expansion, middle, text width=9em, edge=output-black
                [LLM-PKG \cite{wang2024enabling}{,} IKGR \cite{zhengtuning}{,}  CoLaKG \cite{cui2025comprehending}{,} CIKG \cite{hu2025bridging}{,} \\TopicKG \cite{jeon2024topic}{,} LLMRG \cite{wang2023enhancing}{,} ESRS \cite{yong2025enhancing}{,} GSERec \cite{shi2025benefit}{,} \\AutoGraph \cite{shan2024automatic}, leaf, text width=14.5em, edge=output-black]
            ]
        ]
        [Feature Augmentation, middle, text width=7em, edge=output-black
            [GaCLLM \cite{du2024large}{,} HIPHOP\cite{chen2025hierarchical}{,} SKarREC \cite{li2024learning}{,} P4R \cite{chen2024prompting}{,} LIKR \cite{sakurai2024llm}{,} LLaRD \cite{wang2025unleashing}{,} LLMInit \cite{zhang-etal-2025-llminit}{,} \\HGNN4Rec \cite{damianou2024towards}{,} LLM-DMsRec \cite{zhang2025integrating}{,} SEHHN \cite{Xu_Wei_Yang_Wu_2025}{,} DOGE \cite{Meng_Meng_Jin_Lin_Wu_2025}, leaf, text width=25.3em, edge=output-black]
        ]
    ]
    [LLM–Graph Harmonization, middle, edge=output-black, text width=8em
        [Embedding Fusion, middle, text width=7em, edge=output-black
            [LKPNR \cite{runfeng2023lkpnr}{,} DynLLM \cite{zhao2024dynllm}{,} E2EP\cite{ebrat2025end}{,} FreLLM4Rec \cite{wang2025beyond}{,}, leaf, text width=25.3em, edge=output-black]
        ]
        [Embedding Alignment, middle, text width=7em, edge=output-black
            [LLMRec \cite{wei2024llmrec}{,} RLMRec \cite{ren2024representation}{,}  DALR \cite{peng2024denoising}{,} GLTA \cite{yang2025training}{,} KEDRec-LM \cite{zhang2025kedrec}, leaf, text width= 25.3em, edge=output-black]
        ]
    ]
  ]
\end{forest}
\caption{A taxonomy of GFM-based recommender systems.}
\label{fig:taxonomy}
\end{figure*}

From the RS perspective, the recurring need that runs through the entire evolution of recommender systems is not merely "using more features", but using heterogeneous evidence in a way that remains faithful to interaction-driven relevance. Early recommenders learned from co-interactions and IDs; later systems injected metadata and sequences to improve generalization; and modern platforms increasingly demand models that can interpret, reason over, and communicate with rich language content while still exploiting the propagation-like nature of collaborative signals. GFM-style designs can be viewed as a response to this long-standing need: they attempt to reuse the transferable semantic and knowledge capacity of pre-trained models, while re-introducing graph structure as an explicit computational substrate for capturing higher-order collaborative dependencies and relational constraints. At the same time, current GFM-based RS also reveal that "having a powerful LLM" is not automatically sufficient: recommendation relevance is shaped by exposure, feedback sparsity, and platform dynamics, so GFM-based coupling must still be engineered to preserve collaborative evidence and remain efficient at serving scale. Nevertheless, by turning the joint modeling of structure and language into a reusable and extensible recipe, GFM-based RS is increasingly resembling a future paradigm for RS research and practice: a paradigm in which interaction graphs provide the inductive bias for preference propagation, foundation models provide semantic generalization and world knowledge, and the system-level design determines how the two are coordinated to meet accuracy, robustness, and efficiency requirements.

\subsection{Taxonomy of GFM-based RS}
To organize the literature, we categorize GFM-based recommender systems by the dominant coupling direction between graphs and LLMs (Figure~\ref{fig:taxonomy}). Specifically, we group existing methods into three categories:

\begin{itemize}[topsep=0pt]
    \item \textbf{Graph-augmented LLM}, where the structural knowledge from graphs is leveraged to potentiate LLMs' reasoning and generation for recommendation, either implicitly via \emph{token-level} injection into the LLM architecture or explicitly via \emph{context-level} integration into the input prompt;
    \item \textbf{LLM-augmented Graph}, which adopts a data-centric perspective where LLMs serve as powerful synthesizers to refine graph data quality, enhancing either the graph \emph{topology} (e.g., denoising edges) or the node \emph{features} (e.g., enriching textual attributes) to facilitate downstream recommendation tasks;
    \item \textbf{LLM-Graph harmonization}, where the semantic space of LLMs and the structural space of graphs are unified to achieve mutual enhancement, primarily realized through two information flow paradigms: \emph{embedding fusion} for joint modeling and \emph{embedding alignment} for cross-modal consistency.
\end{itemize}

In the following sections, we provide a comprehensive introduction and discussion of the three main categories in the taxonomy of GFM-based RS.

\section{Graph-Augmented LLM}
\label{sec:graph-llm}

LLMs excel at understanding and generating text but struggle with the complex relational structures inherent in recommender systems. While pre-training corpora and in-context learning provide some information, they lack an explicit mechanism to model the intricate relationships between users and items that are naturally represented as graphs.
\begin{figure}[t]
    \centering
    \includegraphics[width=\linewidth]{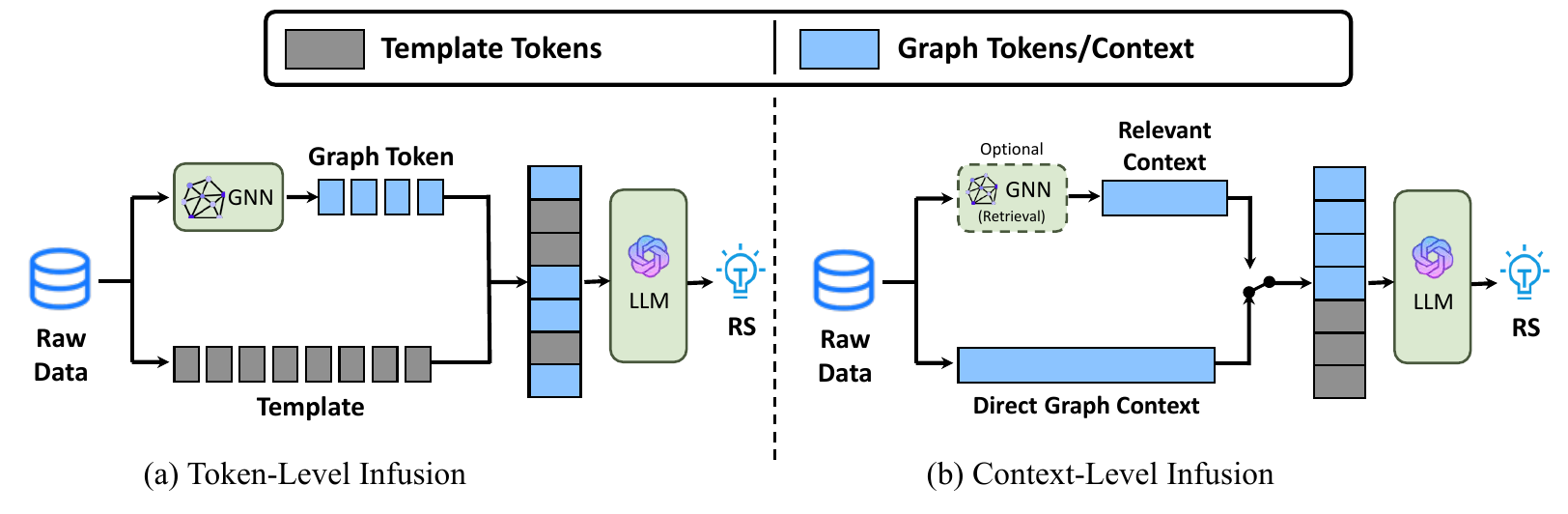}
    \caption{The illustration of graph-augmented LLM methods: 
    a) \textbf{Token-Level Infusion}, where nodes or subgraphs are represented as special tokens, integrating into LLM's input.
    b) \textbf{Context-Level Infusion}, where graph information is converted into context by translating graph into text or retrieving relevant text.}
    \label{fig:graph-augmented LLM}
\end{figure}
Recent research bridges this gap by integrating graphs with LLMs, focusing on the core challenge: \textbf{How to design cross-modal interfaces that effectively bridge graph structures to language models?} We categorize current methods into \textbf{token-level infusion} and \textbf{context-level infusion} (as illustrated in Figure~\ref{fig:graph-augmented LLM}), based on where the cross-modal interface is implemented.

\subsection{Token-Level Infusion}
\label{subsec:token-injection}

This strategy integrates structural information directly into the LLM's input at the token level. Nodes or subgraphs are represented as special tokens, allowing the LLM to process structural information alongside text.

\paragraph{Syntax-Integrated Injection.} This approach embeds special tokens as syntactic components within the LLM's input. For example, TMF~\cite{ma2024triple} introduces \texttt{[ACTION]} tokens like \texttt{[view]} or \texttt{[purchase]} to represent user actions within an interaction sequence. The embeddings for these actions are learned from a multi-behavior graph, enabling the LLM to process semantics like "user \texttt{[views]} item". Building on this, ELMRec~\cite{wang2024enhancing} generates a GCN-based embedding $\mathbf{h}_i$ for each item $i$ and adds it as a correction term to the item's original text embedding $\mathbf{e}_i$, resulting in a refined embedding $\mathbf{e'}_i = \mathbf{e}_i + \mathbf{h}_i$. This operation blends textual and structural information, improving item representation.

A natural progression from here is to consider whether the LLM can directly generates these special tokens. LLMGR~\cite{guo2024integrating} explores this by not only including special tokens in the input but also modifying the LLM's output layer. They introduce a special token for each item, allowing the model to directly generate item tokens as recommendations, effectively creating a tighter link between graph-based recommendations and the LLM's output. Further advancing this line of thought, LightLM~\cite{mei2023lightlm} proposes a hierarchical indexing scheme, which decomposes user/item IDs into multiple special tokens based on a user-item graph-derived index. Each component token encodes a different attribute or function, moving away from opaque numerical IDs towards more semantically meaningful representations.

\paragraph{Syntax-Decoupled Injection.}
This method appends graph embeddings as prefixes or suffixes, separating structural information from the main textual prompt. XRec~\cite{ma2024xrec} prepends GNN-learned embeddings that represent user-item relationships to the prompt. These embeddings are trained to capture high-level semantic concepts, such as preference similarity between users. COMPASS~\cite{qiu2024unveiling} combines user queries with knowledge graph embeddings. A GNN generates embeddings that encapsulate both the user's query and relevant knowledge graph entities, which are then used as a prefix to guide the LLM for recommendation. Optimizing this paradigm for parameter efficiency, Laser~\cite{zhang-etal-2025-bi} introduces a ``Bi-Tuning'' strategy. Instead of retraining the entire model, it injects learnable tokens at both the prefix and suffix of the input sequence. These tokens act as adapters, aligning collaborative signals from the graph with the LLM's semantic space while preserving the model's general capabilities.

Token-level infusion offers a fine-grained way to integrate structural information, allowing for natural interactions between textual and structural data. However, it often requires modifications to the LLM's architecture or careful prompt engineering.

\subsection{Context-Level Infusion}

This strategy provides structural information as context to the LLM, either through text descriptions or implicit retrieval, avoiding modifications to the LLM's architecture.

\paragraph{Explicit Graph-to-Text Mapping.} 
This method involves converting localized graph structures into natural language descriptions, essentially translating graph relationships into text. The following example illustrates explicit graph-to-text mapping: \texttt{user A} $\rightarrow$ \texttt{purchase} $\rightarrow$ \texttt{item B} $\rightarrow$ \texttt{payment} $\rightarrow$ \texttt{credit card} $\Rightarrow$ \texttt{A purchased B with a credit card.}
The simplest form is exemplified by HetGCoT-Rec~\cite{jia2025hetgcot} and KGRec~\cite{abu2024knowledge}, which extract multihop neighbors or relevant entities from knowledge graphs and concatenate their attributes to create natural language descriptions. This explicit serialization provides the LLM with direct factual context. 
Recent advancements have evolved from simple descriptions to more sophisticated semantic alignment and efficiency strategies. K-RagRec~\cite{wang-etal-2025-knowledge-graph} and CORONA~\cite{chen2025corona} refine this process by employing hop-field retrieval and coarse-to-fine chains, respectively, ensuring that only the most semantically relevant subgraphs are serialized into the limited context window.

Beyond factual retrieval, a significant trend focuses on explicitly verbalizing the \textit{reasoning process} inherent in the graph. R2Rec~\cite{zhao2025reason} and CGR~\cite{yang2025personalized} construct explicit ``chains of thought'' derived from graph paths, guiding the LLM to perform certainty-guided reasoning rather than simple pattern matching. Similarly, CausalX~\cite{li2025counterfactual} integrates counterfactual reasoning into the prompt context to mitigate popularity bias, forcing the model to generate explanations based on causal user-item matching.
Furthermore, this explicit mapping has extended to complex and vertical domains. HeLLM~\cite{pv3b-bx85-25} and MSCRS~\cite{wei2025mscrs} tackle the challenge of high-order and multimodal relations by translating hypergraphs and visual-semantic signals into coherent textual prompts. In specialized fields, KERL~\cite{mohbat2025kerl} incorporates domain-specific constraints (e.g., nutritional requirements) into the graph-to-text generation process, ensuring recommendations are both personalized and valid.

\paragraph{Implicit Graph Retrieval.} 
When explicit mapping is difficult or verbose, this approach uses GNN embeddings to retrieve relevant information from the graph semantically. Early works like CLAKG~\cite{chen2024leverage} and URLLM~\cite{shen2024exploring} retrieve legal provisions or historical interactions based on embedding similarity, directly concatenating these records into the prompt. 
However, retrieving raw graph signals can introduce noise. To address this, recent approaches focus on \textit{robustness and denoising}. KERAG\_R~\cite{meng2025kerag_r} employs graph attention mechanisms to filter irrelevant triples before they reach the LLM, while RETURN~\cite{ning2025retrieval} and VoteGCL~\cite{nguyen2025votegcl} utilize collaborative paths and majority voting ensembles to purify adversarial perturbations and augment sparse data, ensuring the retrieved context is high-fidelity.

Another critical direction is optimizing for \textit{dynamics and efficiency}. ColdRAG~\cite{yang2025cold} addresses the cold-start problem by dynamically constructing graphs for new items, allowing immediate retrieval without model retraining. On the inference side, OptiRAG-Rec~\cite{zhou-etal-2025-efficiency} introduces a multi-head early exit mechanism, retrieving graph context only when necessary to balance performance with latency. Finally, in interactive settings, G-Refer~\cite{li2025g} and G-CRS~\cite{qiu2025graph} extend implicit retrieval to conversational agents, dynamically retrieving and translating graph paths to support multi-turn dialogue and explanation generation.

Context-level infusion provides a flexible way to incorporate graph knowledge without altering the LLM's architecture. It leverages the LLM's ability to understand and reason over natural language, making it suitable for scenarios where graph structures can be effectively verbalized.

\subsection{Discussion}
\label{subsec:discussion}

Graph-augmented LLM methods enhance recommendations by encoding rich relational information from graphs, typically through token-level or context-level infusion. This couples the benefits of graph-structured data with the power of LLMs: the graph provides valuable relational context, while the LLM leverages its pre-trained knowledge to interpret it. Furthermore, since the LLM is the central component, this approach augments the recommender system's ability to extrapolate and make inferences in scenarios where interaction data is limited. However, this heavy reliance on the LLM also introduces inherent biases unsuitable for recommendation, such as a lack of diversity and distributional mismatch with user preferences, potentially limiting its scalability and generalizability. The alternative methods, by shifting the focus to enriching or harmonizing the graph itself, effectively mitigate these issues and offer different trade-offs.
\section{LLM-Augmented Graph}
\begin{figure}[t]
    \centering
    \includegraphics[width=\linewidth]{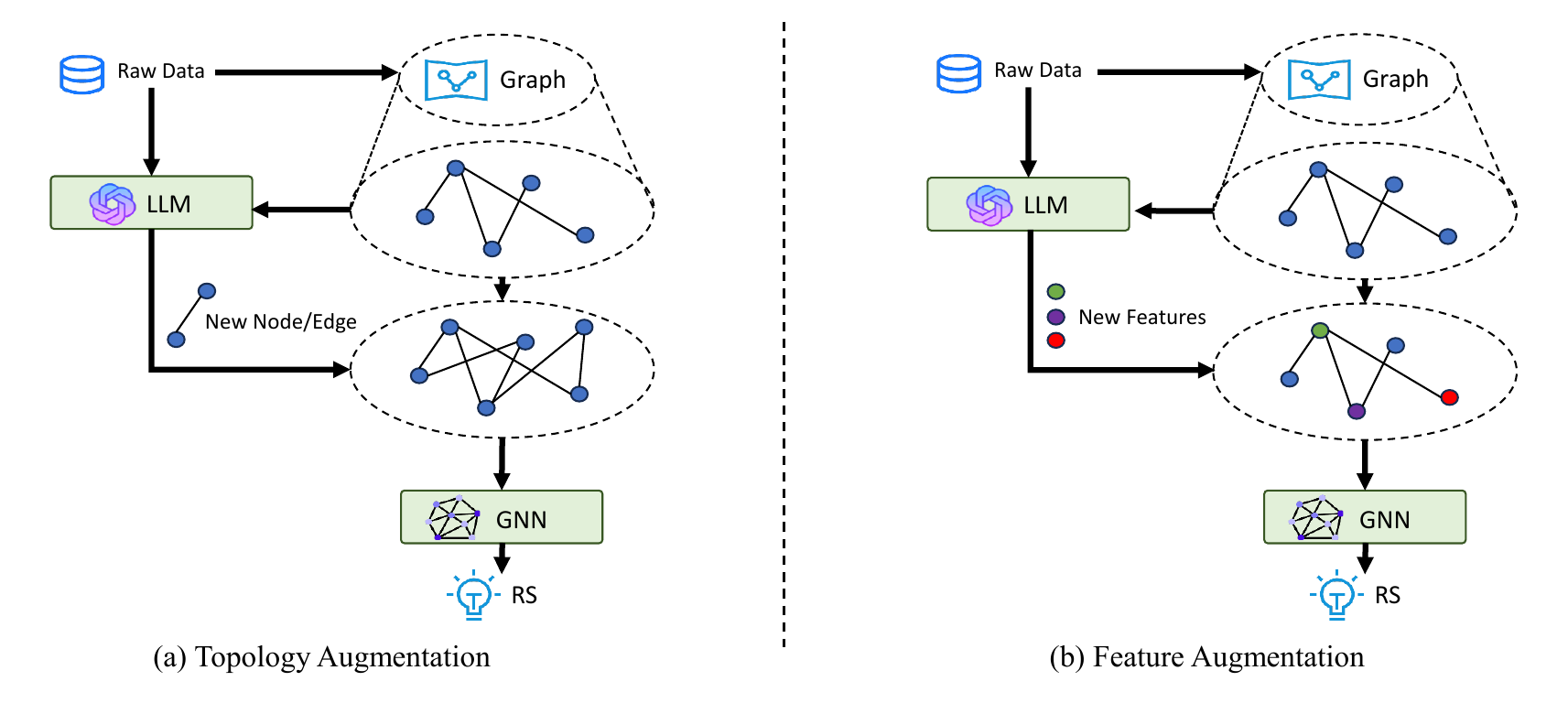}
    \caption{The illustration of LLM-augmented graph methods:  
    a) \textbf{Topology Augmentation}, where LLMs extract structural information from data to alter and augment the topological structure of the graphs.;  
    b) \textbf{Feature Augmentation}, where LLMs processe the textual information in the data, augment the node text or embedding in the graph without changing the topological structure.}
    \label{fig:LLM-augmented graph}
\end{figure}

Shifting the focus to the graph, the core idea of the LLM-augmented graph methods is to augment the data within graphs using LLM, thereby improving the effectiveness of various GNNs employed for recommendation tasks. Such methods can be categorized into \textbf{topology augmentation} and \textbf{feature augmentation} (as illustrated in Figure~\ref{fig:LLM-augmented graph}), based on the aspects of information enhanced in the text-attribute graph according to LLMs.
\subsection{Topology Augmentation}
Topology augmentation refers to the processes where the LLM restructures data, utilizing its world knowledge and contextual understanding capabilities to convert specific textual information into a structured format. Due to the introduction of new structural information, the topological structure of the graph constructed from the data is modified, thereby affecting the subsequent processes to achieve more accurate recommendations.
Intuitively, we categorize topology augmentation into two types: \textbf{edge-level expansion} and \textbf{node-level expansion}, based on whether new nodes are introduced.

\paragraph{Edge-level Expansion.}
This method refers to the process where LLMs introduce new relationships between nodes in the data, such as complementary and substitutable relationships, which are two primary types of relationships of interest in RS.

To directly utilize the versatile capabilities of LLMs and the vast world knowledge, it is intuitive to adopt text-centric approaches for adding edges in the graph. These approaches typically rely on prior knowledge to guide LLMs in making relationship judgments and constructions, either at a superficial or deeper level. As a straightforward example, LLM-KERec~\cite{zhao2024breaking} employs LLMs to assess the complementarity between pairs of items, thereby establishing complementary relationships among item pairs and constructing a complementary item graph. From a deeper perspective, in addition to leveraging complementary relationships between items, subjective user-generated reviews can also be utilized. SAGCN~\cite{liu2023understanding} and FineRec~\cite{zhang2024finerec} utilize LLMs to extract user opinions on items at varying levels of granularity across multiple item attributes (\textit{e.g.}, price, comfort, \textit{etc.}), using this information as edges to construct distinct graphs for each attribute. Beyond complementary or attribute-specific relations, several recent works leverage LLMs to construct more expressive graph structures for recommendation. 
LLM-assisted knowledge graph completion~\cite{abu2025llm} uses LLM reasoning to infer missing relations in domain-specific KGs, thereby enriching graph connectivity and improving downstream recommendation accuracy. 
Confidence-aware KG augmentation~\cite{cai2025boosting} similarly employs LLMs to identify plausible but previously unobserved entity–entity relations, introducing them as new edges while explicitly modeling relation-level confidence to avoid noisy structural expansion. 
Counterfactual reasoning techniques~\cite{li2025counterfactual} further enable LLMs to synthesize hypothetical user--item relations, providing alternative structural views that enhance both recommendation performance and interpretability. 
These approaches demonstrate that LLMs can introduce high-quality structural priors that strengthen graph signals beyond traditional co-occurrence patterns.

Compared to the aforementioned methods, the more refined expansion delves into the relationships within the embedding space, subsequently constructing graphs based on implicit relationships between nodes at a certain level. For example, CSRec~\cite{yang2024common} first employs an LLM to generate complementary or substitutable category nodes based on existing classified nodes, and then utilizes other pre-trained language models to map the pairs of newly generated nodes and existing nodes into the node set within the semantic space. The relationships generated by the LLM are also mapped into the node set within the embedding space. In addition to mapping edges based on semantic similarity, connections can also be established based on the similarity. Several works ~\cite{yang2024sequential,cui2024comprehending} employ an LLM to encode the textual information of nodes into embeddings, and then measure potential relationships between these embeddings through carefully designed methods. These potential relationships serve as edges for item nodes in the graph within the embedding space. Edge expansion has also been explored for improving complementary-product recommendation. 
LLM-enhanced reranking frameworks~\cite{xu2025llm} employ LLMs to evaluate complementary relations between item pairs and incorporate these relations as augmented edges, enabling graph models to better capture cross-item compatibility. 
Such expansions highlight that LLM reasoning is particularly effective for edges that require domain knowledge or semantic coherence.

\paragraph{Node-level Expansion.} 
This method further leverages the world knowledge and contextual reasoning capabilities of LLMs, utilizing auxiliary information as new nodes to augment the information of existing nodes. Such method often directly utilizes the condensed information generated by LLM as new nodes to be introduced into the graph, or extracts the requisite information from the LLM’s output through particular approaches to serve as new nodes in the graph.

Several works~\cite{jeon2024topic,hu2025bridging} directly employ LLMs to generate auxiliary information nodes (\textit{e.g.}, user interests, item categories) for corresponding users or items based on existing textual information. Introducing  such auxiliary information nodes can be viewed as a supplementation of information in the textual space. This approach can, to some extent, assist subsequent GNN in modeling better user or item representations. Furthermore, this type of information supplementation can also be performed in the embedding space. For example, AutoGraph~\cite{shan2024automatic} utilizes LLMs to encode the textual information of users and items, followed by quantizing the semantic embeddings of users and items. By quantizing these semantic embeddings, fine-grained auxiliary information embeddings for users and items can be derived, which are then used as new nodes to supplement information for users and items. Recent progress also shows that LLMs can infer latent intent or semantic factors and introduce them as auxiliary nodes. 
Tuning-Free LLM intent extraction~\cite{zhengtuning} identifies user intent concepts from sparse interaction histories and inserts these inferred intent nodes into the user–item graph to alleviate sparsity. 
Dynamic user knowledge graph construction~\cite{yong2025enhancing} creates new semantic concept nodes representing implicit user preferences to improve serendipity-oriented recommendations. 
GSERec~\cite{shi2025benefit} further introduces discrete semantic code nodes generated via LLM-guided vector quantization, connecting users with learned code representations that act as information bridges in sparse search-enhanced recommendation scenarios. 
Together, these methods demonstrate the increasing importance of introducing LLM-inferred auxiliary nodes as a way to compensate for missing or limited interaction signals.

Topology augmentation represents effectively utilizing the knowledge obtained from LLM pre-training to influence the training of the GNN. The key structural information introduced into the graph during this process enables subsequent GNN to learn more comprehensive representations of users and items.

\subsection{Feature Augmentation}
Enhancing the topological structure of the graph using LLMs may introduce biases, as they are not particularly adept at extracting structural information from text. In contrast, directly improving the node features in the graph without altering the topological structure is an task where LLMs truly excel.
This method focuses on leveraging the natural language processing capabilities of LLMs to augment data at the textual or embedding level, thereby influencing subsequent recommendations.

Some works~\cite{li2024learning,chen2024prompting} enhance the textual information of nodes by constructing appropriate prompts for input into LLMs. The enhanced textual information is subsequently encoded into embeddings by language models such as BERT. Unlike the aforementioned methods, GaCLLM~\cite{du2024large}  integrates the stages of LLM and GNN. In this approach, the LLM assumes the role responsible for message passing within the GNN, performing textual message passing and aggregation for each node in the graph, which is subsequently encoded by BERT. As a special case, LIKR~\cite{sakurai2024llm} employs LLMs to analyze user interaction histories to derive user-preferred item attributes. The nodes corresponding to these attributes in the graph serve as rewards for Markov walk-based reinforcement learning within the graph. A parallel line of work leverages LLMs to enhance node-level semantics without modifying the underlying topology. 
LLM-based denoising frameworks~\cite{wang2025unleashing} extract preference knowledge and relational cues from unstructured interactions to refine node features and reduce noise in collaborative signals. 

The textual information of users and items, being one of the abundant types of information in RS, significantly impacts the performance of RS when effective utilized. Consequently, LLM is increasingly becoming the preferred technology for feature augmentation of graphs based on textual information.

\subsection{Discussion}

The LLM-augmented graph methods, by incorporating the world knowledge of LLM into graph data, can enhance the capabilities of RS at the data level. This makes these methods more competitive in cold start scenarios with sparse interactions. Furthermore, the LLM in these methods is a plug-and-play component, allowing for flexible choices. Moreover, the data processing of the LLM can be performed offline in advance, and online recommendations based on statistical rules or neural models (\textit{e.g.}, GNN) can be made afterwards, significantly reducing time consumption. This has led to the increasing popularity of these methods in industry. However, these methods also have some drawbacks. First, graph learning methods do not fully exploit the world knowledge learned and utilized by LLM, which is a natural shortcoming of such plug-and-play methods. And LLMs have the potential to introduce extraneous noise in both dimensions (topology and feature) of topology augmentation, which may consequently give rise to certain biases. Furthermore, they have poor scalability, as the main body of these methods are shallow GNNs whose model depth is affected by the over-smoothing problem.

\section{LLM-Graph Harmonization}

Graph-augmented LLM methods leverage external graph structures to enhance LLMs but often suffer from inefficiencies in real-time adaptability and increased computational overhead. Conversely, LLM-augmented graph methods attempt to incorporate LLMs into graph-based learning but struggle with scalability and effective knowledge utilization. To address these limitations, this section introduces a novel framework that optimally balances computational efficiency, adaptability, and reasoning capabilities.
\begin{figure}[t]
    \centering
    \includegraphics[width=\linewidth]{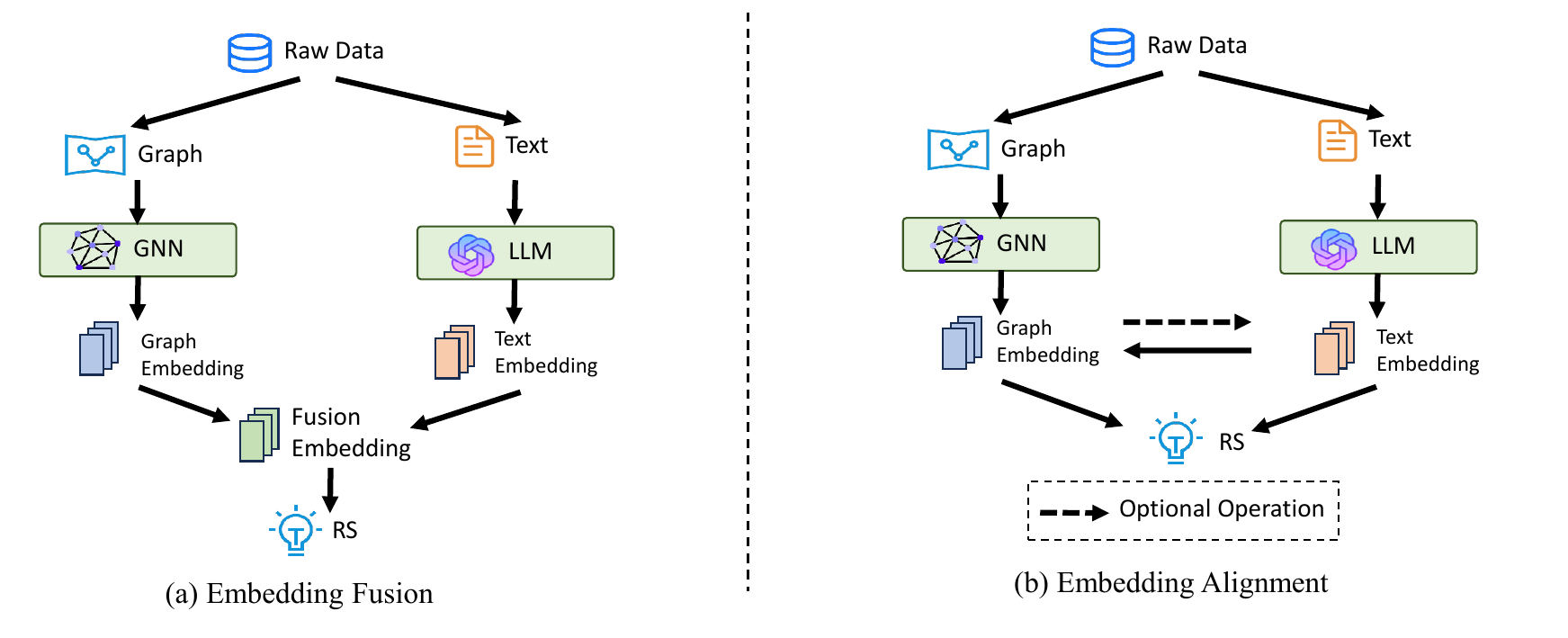}
    \caption{The illustration of LLM-graph harmonization methods:  
    a) \textbf{Embedding Fusion}, where LLM-derived semantic embeddings and GNN-learned structural embeddings are combined into a unified representation space through fusion mechanisms such as concatenation or attention-based integration;  
    b) \textbf{Embedding Alignment}, where embeddings from both modalities are mapped into a shared space using techniques like contrastive learning or MLP-based transformation to enhance consistency and coherence.  }
    \label{fig:LLM-graph harmonization}
\end{figure}

LLMs excel in capturing rich semantic information from unstructured textual data (e.g., item descriptions and user attributes), while GNNs are adept at modeling the topological structure of graphs (e.g., user-item interactions, social connections). As shown in Figure~\ref{fig:LLM-graph harmonization}, harmonizing these two paradigms effectively can significantly enhance recommendation performance. Existing methods can be broadly categorized into two mainstream strategies: \textbf{embedding fusion} and \textbf{embedding alignment}, based on transformations of embeddings.

\subsection{Embedding Fusion}

The embedding fusion approach aims to combine LLM-derived textual representations with graph-learned structural embeddings, creating a unified feature space that leverages complementary information. This strategy emphasizes the synergy between textual semantics and graph-based connectivity.


A notable framework in this domain is DynLLM~\cite{zhao2024dynllm}, which incorporates graph structures into LLMs through dynamic memory-enhanced fusion. DynLLM addresses the limitation of static embeddings by using a dual-flow interaction mechanism: one flow learns from GNN-updated dynamic embeddings reflecting user-item interactions, while the other adapts LLM-generated embeddings based on real-time textual content. These embeddings are fused in a shared latent space, enhancing both semantic and structural understanding. Such a dynamic fusion approach not only captures the temporal evolution of recommendation data but also integrates high-quality textual semantics, leading to more accurate and adaptive recommendations. Another example is LKPNR~\cite{runfeng2023lkpnr}, which combines LLMs with knowledge graphs to enhance personalized news recommendation. LKPNR leverages the semantic richness of LLMs to generate high-quality news representations and uses knowledge graphs to capture the relational structure of news entities. By integrating these modalities, LKPNR effectively addresses the long-tail problem in news recommendation. In addition to these representative approaches, several recent works further extend the fusion paradigm by enabling deeper semantic--structural integration. 
End-to-End Personalization~\cite{ebrat2025end} unifies LLM-based semantic modeling and graph-based interaction modeling within a single optimization framework, effectively fusing LLM-generated representations with collaborative structural embeddings for holistic personalization. 
Beyond Semantic Understanding~\cite{wang2025beyond} highlights that LLM-based recommenders often underutilize collaborative frequency signals; the authors propose a fusion mechanism that preserves frequency-domain interaction components alongside semantic embeddings. 
These advances demonstrate a growing trend toward fusion models that jointly encode both semantic richness and collaborative patterns, thereby improving robustness and expressiveness in large-scale recommenders.

The embedding fusion paradigm is particularly effective because it exploits the contextual richness of LLMs alongside the relational structures captured by GNNs. By fusing these modalities, models like DynLLM and LKPNR enable direct, dynamic, and efficient utilization of both textual and graph data, significantly improving representation learning in recommendation scenarios.

\subsection{Embedding Alignment}

Embedding alignment takes a different route by focusing on reconciling the heterogeneity between LLM-generated textual embeddings and GNN-learned structural representations. This strategy ensures that embeddings from both modalities can operate coherently within a unified representational space, reducing information loss and noise. The refined embeddings can provide more valuable information for recommendations.


The DALR framework~\cite{peng2024denoising} serves as a representative example, where structural embeddings from GNNs (e.g., user-item graph representations) and semantic embeddings from LLMs (e.g., product descriptions, user reviews) are aligned through contrastive learning paradigms. This alignment mitigates the semantic gaps and noise introduced by the inherently different data sources. Similarly, methods such as LLMRec~\cite{wei2024llmrec} and RLMRec~\cite{ren2024representation} also adopt multimodal alignment techniques, such as contrastive learning and MLP-based alignment, to unify embeddings from diverse modalities. For instance, LLMRec employs a denoised data robustification mechanism to enhance the reliability of augmented recommendation data, while RLMRec leverages contrastive alignment strategies to bridge the semantic space of LLMs with the collaborative relational signals from GNNs, thereby improving the overall quality of the learned representations. Recently, several works have pushed alignment-based harmonization toward more explicit cross-modal consistency. 
Graph-Language Token Alignment (GLTA)~\cite{yang2025training} constructs a shared token space by aligning graph nodes with LLM token embeddings, enabling structural relationships to be directly encoded within the LLM vocabulary. 
Similarly, KEDRec-LM~\cite{zhang2025kedrec} adopts a knowledge-distillation-based alignment strategy, transferring domain-specific knowledge graph representations into the LLM embedding space to improve interpretability and accuracy in medical recommendation. 
These approaches highlight the versatility of embedding alignment: rather than fusing modalities, they enforce representational coherence between graph encoders and LLMs, making them especially effective in knowledge-intensive recommendation scenarios.

Embedding alignment excels in its ability to unify heterogeneous modalities, fostering a coherent embedding space that captures complex inter-modal dependencies. By ensuring consistent representation learning, it provides a robust foundation for downstream recommendation tasks, ranging from fine-grained personalized ranking to latent user preference clustering, thereby enhancing the model's sensitivity to multi-faceted user intents.

\subsection{Discussion}

The two approaches offer distinct advantages in integrating LLMs with graph learning for RS. Embedding fusion directly combines textual semantics and structural relationships, leveraging their complementarity to enhance representation learning and personalization. Dynamic fusion methods, such as DynLLM, further enable real-time adaptation to evolving user preferences. Embedding alignment, in contrast, ensures coherence between textual and structural embeddings by mapping them into a shared space, mitigating inconsistencies. Methods like DALR leverages various contrastive learning approaches to enhance alignment robustness. However, embedding fusion may introduce redundant or conflicting information, and increase computational costs, while embedding alignment which is sensitive to noise depends on high-quality training data.
In these methods, the LLM primarily serves as an encoder, and due to the constraints of the scenario, it cannot fully utilize its contextual understanding and language generation capabilities.

{\centering
\small
\renewcommand{\arraystretch}{1} 

\begin{longtable}{p{2.5cm}lp{8cm}} 

\caption{Different types of datasets used in GFM-based RS} \label{tab:datasets} \\
\toprule
\textbf{Categories} & \textbf{Datasets} & \textbf{Reference} \\
\midrule
\endfirsthead

\multicolumn{3}{c}%
{{\bfseries \tablename\ \thetable{} -- continued from previous page}} \\
\toprule
\textbf{Categories} & \textbf{Datasets} & \textbf{Reference} \\
\midrule
\endhead

\bottomrule
\endfoot

\multirow{5}{2.5cm}{Interaction-Centric}
 & Netflix & CORONA~\cite{chen2025corona}, VoteGCL~\cite{nguyen2025votegcl}, LLMRec~\cite{wei2024llmrec} \\ \cmidrule{2-3}
 & MovieLens & CORONA~\cite{chen2025corona}, CoLaKG~\cite{cui2025comprehending}, E2EP~\cite{ebrat2025end}, CIKG~\cite{hu2025bridging}, CausalX~\cite{li2025counterfactual}, KERAG\_R~\cite{meng2025kerag_r}, VoteGCL~\cite{nguyen2025votegcl}, RETURN~\cite{ning2025retrieval}, LIKR~\cite{sakurai2024llm}, AutoGraph~\cite{shan2024automatic}, K-RagRec~\cite{wang-etal-2025-knowledge-graph}, LLMRec~\cite{wei2024llmrec}, GLTA~\cite{yang2025training}, LRD~\cite{yang2024sequential}, LLM-DMsRec~\cite{zhang2025integrating}, R2Rec~\cite{zhao2025reason} \\ \cmidrule{2-3}
 & Last-FM & CoLaKG~\cite{cui2025comprehending}, RETURN~\cite{ning2025retrieval}, LIKR~\cite{sakurai2024llm}, FreLLM4Rec~\cite{wang2025beyond} \\
 \cmidrule{2-3}
 & Diginetica & HIPHOP~\cite{chen2025hierarchical} \\ \cmidrule{2-3}
 & Yoochoose & HIPHOP~\cite{chen2025hierarchical} \\ 
\midrule

\multirow{8}{2.5cm}{Semantic-Rich Textual}
 & Yelp & GAL-Rec~\cite{guan2024enhancing}, G-Refer~\cite{li2025g}, HGPO~\cite{luo2025llm}, LightLM~\cite{mei2023lightlm}, XRec~\cite{ma2024xrec}, VoteGCL~\cite{nguyen2025votegcl}, DALR~\cite{peng2024denoising}, RLMRec~\cite{ren2024representation}, LLaRD~\cite{wang2025unleashing}, SEHHN~\cite{Xu_Wei_Yang_Wu_2025}, FineRec~\cite{zhang2024finerec}, IKGR~\cite{zhengtuning} \\ \cmidrule{2-3}
 & Amazon & CKG-LLMA~\cite{cai2025boosting}, HIPHOP \cite{chen2025hierarchical}, P4R~\cite{chen2024prompting}, CORONA~\cite{chen2025corona}, GAL-Rec~\cite{guan2024enhancing}, LLMGR~\cite{guo2024integrating}, HeLLM~\cite{pv3b-bx85-25}, TopicKG~\cite{jeon2024topic}, CausalX~\cite{li2025counterfactual}, G-Refer~\cite{li2025g}, SAGCN~\cite{liu2023understanding}, HGPO~\cite{luo2025llm}, LightLM~\cite{mei2023lightlm}, DOGE~\cite{Meng_Meng_Jin_Lin_Wu_2025}, XRec~\cite{ma2024xrec}, KERAG\_R~\cite{meng2025kerag_r}, DALR~\cite{peng2024denoising}, RLMRec~\cite{ren2024representation}, URLLM~\cite{shen2024exploring}, AutoGraph~\cite{shan2024automatic}, TWISTER~\cite{wang2025towards}, FreLLM4Rec~\cite{wang2025beyond}, K-RagRec~\cite{wang-etal-2025-knowledge-graph}, LLaRD~\cite{wang2025unleashing}, ELMRec~\cite{wang2024enhancing}, LLaRD~\cite{wang2025unleashing}, SEHHN~\cite{Xu_Wei_Yang_Wu_2025}, LLM-ER~\cite{xu2025llm}, GLTA~\cite{yang2025training}, CSRec~\cite{yang2024common}, LRD~\cite{yang2024sequential}, ColdRAG~\cite{yang2025cold}, TagCF~\cite{yuyou}, LLM-DMsRec~\cite{zhang2025integrating}, LLMInit~\cite{zhang-etal-2025-llminit}, Laser~\cite{zhang-etal-2025-bi}, FineRec~\cite{zhang2024finerec}, R2Rec~\cite{zhao2025reason}, IKGR~\cite{zhengtuning}, OptiRAG-Rec~\cite{zhou-etal-2025-efficiency} \\ \cmidrule{2-3}
 & Google reviews & G-Refer~\cite{li2025g}, XRec~\cite{ma2024xrec}, VoteGCL~\cite{nguyen2025votegcl} \\ \cmidrule{2-3}
 & MIND & CoLaKG~\cite{cui2025comprehending}, HGPO~\cite{luo2025llm}, LKPNR~\cite{runfeng2023lkpnr} \\ \cmidrule{2-3}
 & Goodreads & TWISTER~\cite{wang2025towards}, GLTA~\cite{yang2025training} \\ \cmidrule{2-3}
 & Steam & CKG-LLMA~\cite{cai2025boosting}, DALR~\cite{peng2024denoising}, RLMRec~\cite{ren2024representation}, LLaRD~\cite{wang2025unleashing}, SEHHN~\cite{Xu_Wei_Yang_Wu_2025}, IKGR~\cite{zhengtuning} \\ \cmidrule{2-3}
 & Book-Crossing & CIKG~\cite{hu2025bridging}, AutoGraph~\cite{shan2024automatic}, OptiRAG-Rec~\cite{zhou-etal-2025-efficiency} \\ \cmidrule{2-3}
 & Anime & CKG-LLMA~\cite{cai2025boosting} \\
\midrule

\multirow{5}{2.5cm}{Knowledge-Driven Structured}
 & DBbook & CIKG~\cite{hu2025bridging} \\ \cmidrule{2-3}
 & DBLP & HetGCoT-Rec~\cite{jia2025hetgcot} \\ \cmidrule{2-3}
 & OpenAlex & HetGCoT-Rec~\cite{jia2025hetgcot} \\ \cmidrule{2-3}
 & Junyi & SKarREC~\cite{li2024learning} \\ \cmidrule{2-3}
 & ASSISTments & SKarREC~\cite{li2024learning} \\
\midrule

\multirow{2}{2.5cm}{Conversational Dialogue}
 & ReDial & COMPASS~\cite{qiu2024unveiling}, G-CRS~\cite{qiu2025graph}, MSCRS~\cite{wei2025mscrs} \\ \cmidrule{2-3}
 & INSPIRED & COMPASS~\cite{qiu2024unveiling}, G-CRS~\cite{qiu2025graph}, MSCRS~\cite{wei2025mscrs} \\

\end{longtable}}

\section{Datasets for GFM-based RS}
\label{sec:datasets}

GFM-based recommender systems are motivated by a practical observation: recommendation data often contains \emph{relational structure} (e.g., interaction graphs, session transitions, heterogeneous relations, or knowledge triples) and \emph{natural-language semantics} (e.g., reviews, descriptions, tags, or dialogues), yet these signals are usually exploited by separate modeling paradigms.
Consequently, datasets used in this area vary not only by scale, but also by what they make \emph{easy} or \emph{hard}: some emphasize pure collaborative structure with minimal text, while others provide dense language that often needs to be grounded in sparse or noisy interactions; still others introduce explicit multi-relational structure (knowledge graphs, concept hierarchies, scholarly networks) or multi-turn conversational contexts.

To make these differences explicit for evaluation and reproducibility, we organize commonly used datasets in GFM-based RS into four groups according to the \emph{dominant} data form in typical pipelines: (i) \textbf{Interaction-Centric} datasets that primarily benchmark collaborative/sequential modeling from interaction graphs, (ii) \textbf{Semantic-Rich Textual} datasets that pair interactions with substantial text, (iii) \textbf{Knowledge-Driven Structured} datasets that emphasize explicit multi-relational structure (e.g., knowledge graphs, heterogeneity, or concept dependencies), and (iv) \textbf{Conversational Dialogue} datasets that frame recommendation as multi-turn natural-language interaction.
These groups are not strictly mutually exclusive—many datasets contain multiple signal sources—but they reflect how datasets are most frequently used in existing GFM-based RS studies.

\subsection{Interaction-Centric Datasets}
\label{subsec:datasets:interaction}

Interaction-centric datasets provide large-scale user behavior logs with limited textual content, and are therefore a clean testbed for assessing whether a GFM can improve over strong graph-based baselines when the semantic channel is weak.
They are also where graph inductive bias matters most: modeling sparsity, long-tail distributions, temporal dynamics, and higher-order connectivity typically dominates performance.
In practice, these datasets are frequently instantiated as a user--item bipartite graph (for long-term preference modeling) or as a session graph (for short-term intent modeling).

\paragraph{Netflix.}
Netflix is a canonical benchmark for collaborative filtering, consisting primarily of explicit ratings with rating dates and minimal textual content.
Because it contains little usable natural language for most setups, it is commonly used to evaluate the ``structural'' side of GFM-based RS: whether the model can preserve and exploit interaction topology (including long-range collaborative dependencies) without relying on rich content features.
For GFM pipelines, Netflix is also useful to test whether language modules (e.g., instruction-tuned components or reasoning-oriented heads) can still contribute via better training objectives, calibration, or retrieval-style scaffolding even when item semantics are not directly provided.

\paragraph{MovieLens.}
MovieLens remains the most widely adopted academic dataset family for rating-based recommendation.
Beyond explicit feedback and timestamps, many versions include lightweight semantics such as user-generated tags, which are small compared to review corpora but still helpful for semantic alignment.
This makes MovieLens an attractive ``controlled'' benchmark: researchers can vary the amount of semantics (ratings only vs.\ ratings+tags) and examine how GFM components shift the balance between graph propagation and semantic priors.

\paragraph{Last-FM.}
Last.fm datasets are representative of implicit-feedback recommendation, often providing abundant timestamped listening events.
Depending on the release and preprocessing, they can include tags (e.g., artist or track tags) and user/item attributes, enabling both sequential preference modeling and limited semantic grounding.
For GFMs, Last-FM is particularly useful when studying popularity bias, long-tail preference distribution, and temporal regularities: the model typically needs to represent fine-grained user taste from repeated events while remaining robust to sparse side information.

\paragraph{Diginetica.}
Diginetica is a standard session-based recommendation benchmark, where the primary signal is short click/purchase sequences rather than persistent user profiles.
It is commonly transformed into session graphs (items as nodes, transitions as directed edges) to capture within-session dependencies.
From the GFM perspective, Diginetica stresses short-horizon intent understanding under extremely limited context, which is a different regime from long-term preference graphs and often exposes whether a model can leverage lightweight semantics or generalization priors to stabilize predictions.

\paragraph{Yoochoose.}
Yoochoose (RecSys Challenge 2015) is another classic session-based dataset, notable for its scale and for the fact that sessions are often anonymous.
Similar to Diginetica, it is typically modeled as a session graph or a sequence graph for next-item prediction, emphasizing scalability and robustness under highly skewed item popularity.
For GFM-based approaches, Yoochoose is valuable because it separates ``modeling structure'' from ``understanding content'': gains typically come primarily from how the model represents transitions, context windows, and higher-order session patterns, rather than from rich text.

\subsection{Semantic-Rich Textual Datasets}
\label{subsec:datasets:text}

Semantic-rich textual datasets explicitly couple interaction graphs with abundant natural-language content such as reviews, item descriptions, titles, and tags.
They are the most direct benchmarks for GFM-based RS, because they involve non-trivial \emph{grounding}: text is informative but noisy, and user-generated language does not always correspond cleanly to preference strength or ranking behavior.
These datasets are therefore often used to study cold-start and long-tail generalization, robust alignment between semantic and collaborative representations, and explanation- or reasoning-aware recommendation.

\paragraph{Yelp.}
Yelp datasets are commonly used as heterogeneous recommendation benchmarks, where the core structure is a user--business interaction graph (often with explicit ratings) and the semantic channel is review text.
In many setups, business metadata (e.g., categories and attributes) further enriches node features, making it possible to study multi-source fusion rather than review-only modeling.
For GFMs, Yelp is particularly suitable for evaluating whether review semantics (sentiment, aspects, and fine-grained opinions) can be aligned with graph-derived preference signals, and whether the resulting representations improve robustness in sparse neighborhoods.

\paragraph{Amazon.}
The Amazon Reviews datasets constitute a large family of domain-specific corpora that combine interactions (ratings and/or implicit feedback), rich review text, and multi-field item metadata (titles, categories, brand, etc.).
Some releases and derived processing pipelines additionally provide item-to-item relations (e.g., co-purchase/co-view), which can naturally induce multi-relational graphs beyond a plain user--item bipartite structure.
This combination makes Amazon a versatile benchmark for GFM-based RS: models often need to handle extreme sparsity, domain shift across categories, and the frequent mismatch between textual verbosity and true preference intensity.

\paragraph{MIND.}
MIND is tailored for news recommendation, pairing impression/click logs with rich textual fields such as titles and abstracts (and potentially other article metadata).
Compared with review-based datasets, its defining characteristic is fast temporal drift: user interests can change quickly, and items (news articles) have short lifecycles.
GFM-based RS evaluated on MIND therefore often needs to reconcile dynamic interaction structure with high-entropy text, and this setting is well-suited for analyzing whether graph--language coupling improves time-sensitive generalization rather than merely memorizing item semantics.

\paragraph{Goodreads.}
Goodreads datasets typically include user--book feedback alongside abundant user-generated content, such as reviews and tags/shelves.
In contrast to Yelp (often aspect-centric) and Amazon (often category-centric), Goodreads text tends to be longer-form and can reflect subjective narratives, which introduces additional noise and abstraction.
This makes it a useful benchmark for GFM alignment: the model can learn to distill preference-relevant semantics from lengthy text while still grounding recommendations in collaborative structure.

\paragraph{Steam.}
Steam datasets combine user--game interactions with review text and, in many curated versions, additional behavioral signals such as playtime.
The coexistence of multiple engagement signals is important: playtime can indicate interest but may diverge from explicit satisfaction, and reviews may be biased toward extreme opinions.
For GFMs, Steam is a strong testbed for multi-signal representation learning and for studying whether language semantics can help explain or refine behavior-only inference.

\paragraph{Book-Crossing.}
Book-Crossing provides large-scale user--book interactions and \emph{bibliographic metadata} (e.g., title/author/publisher/year), which serves as structured semantic context even when free-form reviews are limited.
It is frequently used in sparse recommendation settings, where semantic side information can be crucial for representing cold-start books.
For GFM-based RS, Book-Crossing is valuable for exploring how language-capable components leverage short textual fields and structured attributes, rather than long reviews.

\paragraph{Anime.}
Anime datasets typically contain explicit ratings and a set of structured content attributes (genres, types, episodes, etc.), sometimes with brief descriptions depending on the source.
Although the semantics here are more categorical than textual, they still provide a meaningful semantic channel that can be mapped into language or embedding space.
This dataset is often used to study preference clustering and fine-grained attribute alignment, where an important question is whether GFM coupling improves representation beyond standard graph+feature baselines.

\paragraph{Google reviews.}
Google review corpora typically contain user ratings and free-text reviews for \emph{points of interest (POIs)} (i.e., physical places such as restaurants, hotels, shops, and attractions) along with auxiliary fields like business categories and coarse-grained location information (depending on the release).
Compared with product recommendation, POI recommendation is often more sensitive to contextual factors: user preferences may correlate with geography, and popularity can be strongly shaped by local exposure and tourist traffic.
For GFM-based RS, these datasets are useful for studying whether a model can ground review semantics (opinions, aspects, situational cues) in user--place interaction graphs, while remaining robust to location- and popularity-driven confounders.

\subsection{Knowledge-Driven Structured Datasets}
\label{subsec:datasets:kg}

Knowledge-driven structured datasets emphasize explicit relational structure beyond simple user--item interactions, including knowledge graphs, heterogeneous schemas, or concept hierarchies.
They naturally support evaluation of relation-aware message passing, multi-hop reasoning, and semantic grounding through entity and concept descriptions.
In GFM-based RS, these datasets are frequently used to test whether language models can help interpret, augment, or align structured relations—while graph encoders preserve compositional structure that is hard to express purely as token sequences.

\paragraph{DBbook.}
DBbook is a sparse rating dataset where items can be linked to external semantic resources (e.g., via DBpedia URIs), enabling the inclusion of knowledge-graph attributes and relations.
Its sparsity highlights a core motivation for GFMs: when interaction evidence is weak, structured knowledge and semantics can provide crucial inductive bias.
Accordingly, DBbook is often used to study cold-start robustness and the effectiveness of knowledge-grounded representation learning.

\paragraph{DBLP.}
DBLP-based corpora form heterogeneous scholarly graphs connecting authors, papers, venues, years, and citation/co-authorship relations, frequently with paper titles as textual fields.
Unlike standard RS datasets, the ``user'' and ``item'' roles can be task-dependent (e.g., author--paper, paper--paper), which makes the schema more complex and closer to real-world knowledge-intensive recommendation.
For GFMs, DBLP highlights the need to jointly model heterogeneous topology and scientific language semantics under multi-relational dependencies.

\paragraph{OpenAlex.}
OpenAlex provides a large-scale scholarly knowledge graph with rich entities (works, authors, institutions, concepts) and explicit relations.
Its scale and heterogeneity make it attractive for stress-testing GFM scalability and retrieval-style augmentation, especially when tasks require reasoning across multiple entity types (e.g., concept-aware paper recommendation or expert finding).
It also offers an environment where language signals (e.g., titles and other available textual metadata, depending on preprocessing) can be integrated with complex structure rather than a single bipartite graph.

\paragraph{Junyi.}
Junyi is an education-domain dataset centered on student--problem interaction sequences, often accompanied by knowledge components, skill tags, or prerequisite-like structure.
It supports evaluation of educational recommendation and knowledge tracing, where dependencies between concepts matter.
For GFM-based RS, Junyi is useful to examine whether graph--language coupling can model structured concept relations while capturing temporal learning dynamics.

\paragraph{ASSISTments.}
ASSISTments datasets record student responses to exercises linked to knowledge concepts, producing a multi-entity structure (student, problem, concept) and sequential logs.
This setting stresses structured dependency modeling under temporal sequences: the model typically needs to represent how mastery evolves and how concepts relate.
GFMs can be evaluated here for their ability to align concept-level structure (graph) with semantic descriptors (concept names, exercise content when available) and to provide more generalizable student representations.

\subsection{Conversational Dialogue Datasets}
\label{subsec:datasets:dialogue}

Conversational dialogue datasets formulate recommendation as multi-turn interaction, where user intent is expressed in natural language and the system needs to decide what to recommend (and how to respond) given evolving dialogue context.
These benchmarks are important for evaluating whether GFMs can unify dialogue understanding, preference elicitation, and graph-based grounding over items/entities.

\paragraph{ReDial.}
ReDial consists of human--human conversations about movie recommendation, and is commonly associated with an item universe that can be linked to external knowledge bases.
It captures preference revelation through dialogue acts and natural-language mentions rather than static profiles.
For GFM-based conversational RS, ReDial is valuable because it typically calls for both semantic understanding of dialogue context and structured grounding over items/entities to produce coherent, relevant recommendations.

\paragraph{INSPIRED.}
INSPIRED contains recommendation dialogues with annotations emphasizing social and conversational strategies in addition to task success.
Compared with purely information-seeking dialogues, it makes social behavior an explicit part of the evaluation context.
This dataset is therefore well-suited for studying GFM-based models that need to balance recommendation grounding (via graphs/knowledge) with natural and socially appropriate response generation.

\section{Challenges and Future Directions}

GFMs have demonstrated great potential in recommender systems by incorporating graph structural information  with external world knowledge of LLM. However, several challenges hinder their widespread adoption and effectiveness.

\paragraph{High Computational Cost and Scalability Issues.}
Existing GFM-based RS require substantial computational resources, posing challenges for large-scale deployment \cite{zhai2024actions}. The integration of graph-based reasoning and LLM inference results in high memory consumption and slow inference speed, particularly when processing dense user-item graphs or generating personalized recommendations in real-time \cite{10.1145/3640457.3688161}. Unlike traditional recommendation models, which can be efficiently pruned or quantized, GFMs face unique scalability constraints due to their reliance on long-range graph dependencies and LLM-generated representations. Addressing these limitations requires advancements in efficient model compression strategies tailored for graph-enhanced LLMs, and adaptive graph sparsification techniques to maintain performance while reducing overhead.

\paragraph{Robustness Against Noisy and Adversarial Data.}
Real-world user interactions are inherently noisy, exhibiting short-term fluctuations, incomplete preferences, and adversarial perturbations \cite{zhang2023robust}. Traditional recommendation models rely on explicit feedback signals, making them susceptible to biased or manipulated data. In contrast, GFMs integrate graph-based user-item relationships and LLM-generated contextual representations, which introduces additional sources of noise from both structured and unstructured data. Ensuring robustness requires advancements in self-supervised denoising techniques, adversarial training tailored for multimodal representations, and uncertainty-aware modeling to mitigate the impact of unreliable signals while preserving recommendation accuracy.

\paragraph{Multi-Modal Information Fusion.}
Modern recommendation scenarios involve a diverse range of data modalities, including text, structured graphs, images, audio, and video \cite{tao2020mgat}. While existing GFMs primarily focus on textual and structural embeddings, effectively incorporating rich multi-modal signals remains an open challenge. Different modalities exhibit varying levels of granularity, semantic gaps, and computational costs, making seamless integration nontrivial. Future research should explore adaptive fusion frameworks, cross-modal alignment mechanisms, and lightweight multi-modal representation learning to balance efficiency and accuracy in large-scale recommender systems.

\paragraph{Lack of End-to-End Optimization.}  
The concept of end-to-end recommender system is not unfamiliar. When deep learning was introduced into the field of recommendation, a process encapsulation was essentially performed~\cite{covington2016deep}. However, the early neural model RS have gradually fallen behind the times. The process of such RS can be roughly divided into three stages: matching, ranking, and re-ranking~\cite{gao2023survey}. In reality, this process is often more refined in industrial applications. Such a meticulous process naturally results in better recommendation performance. However, multi-stage model optimization requires a significant investment of time and manpower. Contrarily, an end-to-end generative RS, different from the one mentioned above, encapsulates multiple stages together for optimization. This significantly reduces complexity and can potentially lead to better performance. Gradually, similar endeavors are being pursued in the industrial field. HSTU~\cite{zhai2024actions} simplifies the internal structure of the LLM and fully implements it through serial modeling. Moreover, \cite{wang2024llm} takes into account both structural and textual information. Such an integrated generative recommendation that combines matching and ranking may likely be a hotspot in the future.

\paragraph{Knowledge-Preference Gap.}
While GFMs leverage external knowledge to alleviate data sparsity, a fundamental misalignment persists between globally pre-trained world knowledge and personalized user preferences \cite{10.1145/3640457.3688161}. Unlike embedding alignment, which focuses on bridging modality gaps (e.g., between graph structures and textual representations), this discrepancy stems from differences in how LLMs interpret knowledge and how users express preferences. For instance, LLM-based recommendation models may naturally generate factually coherent but overly neutral item descriptions, whereas users often respond more favorably to engaging or sensationalized content (e.g., ``Shocking! You won’t believe this..."). Addressing this challenge requires advancing preference-aware knowledge adaptation, dynamic refinement techniques, and contrastive learning strategies tailored to user-specific interests.

\section{Conclusion}
As an indispensable technology in modern society, recommender systems stand as one of the most prominent research areas within the field of artificial intelligence. The emergence of graph foundation models is likely to spark a new wave of research enthusiasm in the field of RS. In this survey, we present the first comprehensive overview of GFM-based RS and propose a logically organized taxonomy. Furthermore, we delve deeply into the challenges and vast potential of this field, aiming to inject new vitality into its research endeavors.


\bibliographystyle{ACM-Reference-Format}
\bibliography{ijcai25}


\begin{thebibliography}{111}


\ifx \showCODEN    \undefined \def \showCODEN     #1{\unskip}     \fi
\ifx \showDOI      \undefined \def \showDOI       #1{#1}\fi
\ifx \showISBNx    \undefined \def \showISBNx     #1{\unskip}     \fi
\ifx \showISBNxiii \undefined \def \showISBNxiii  #1{\unskip}     \fi
\ifx \showISSN     \undefined \def \showISSN      #1{\unskip}     \fi
\ifx \showLCCN     \undefined \def \showLCCN      #1{\unskip}     \fi
\ifx \shownote     \undefined \def \shownote      #1{#1}          \fi
\ifx \showarticletitle \undefined \def \showarticletitle #1{#1}   \fi
\ifx \showURL      \undefined \def \showURL       {\relax}        \fi
\providecommand\bibfield[2]{#2}
\providecommand\bibinfo[2]{#2}
\providecommand\natexlab[1]{#1}
\providecommand\showeprint[2][]{arXiv:#2}

\bibitem[Abu-Rasheed et~al\mbox{.}(2025)]%
        {abu2025llm}
\bibfield{author}{\bibinfo{person}{Hasan Abu-Rasheed}, \bibinfo{person}{Constance Jumbo}, \bibinfo{person}{Rashed Al~Amin}, \bibinfo{person}{Christian Weber}, \bibinfo{person}{Veit Wiese}, \bibinfo{person}{Roman Obermaisser}, {and} \bibinfo{person}{Madjid Fathi}.} \bibinfo{year}{2025}\natexlab{}.
\newblock \showarticletitle{Llm-assisted knowledge graph completion for curriculum and domain modelling in personalized higher education recommendations}. In \bibinfo{booktitle}{\emph{2025 IEEE Global Engineering Education Conference (EDUCON)}}. IEEE, \bibinfo{pages}{1--5}.
\newblock


\bibitem[Abu-Rasheed et~al\mbox{.}(2024)]%
        {abu2024knowledge}
\bibfield{author}{\bibinfo{person}{Hasan Abu-Rasheed}, \bibinfo{person}{Christian Weber}, {and} \bibinfo{person}{Madjid Fathi}.} \bibinfo{year}{2024}\natexlab{}.
\newblock \showarticletitle{Knowledge Graphs as Context Sources for LLM-Based Explanations of Learning Recommendations}.
\newblock \bibinfo{journal}{\emph{arXiv preprint arXiv:2403.03008}} (\bibinfo{year}{2024}).
\newblock


\bibitem[Adomavicius and Tuzhilin(2005)]%
        {adomavicius2005toward}
\bibfield{author}{\bibinfo{person}{Gediminas Adomavicius} {and} \bibinfo{person}{Alexander Tuzhilin}.} \bibinfo{year}{2005}\natexlab{}.
\newblock \showarticletitle{Toward the next generation of recommender systems: A survey of the state-of-the-art and possible extensions}.
\newblock \bibinfo{journal}{\emph{IEEE transactions on knowledge and data engineering}} \bibinfo{volume}{17}, \bibinfo{number}{6} (\bibinfo{year}{2005}), \bibinfo{pages}{734--749}.
\newblock


\bibitem[Baluja et~al\mbox{.}(2008)]%
        {baluja2008video}
\bibfield{author}{\bibinfo{person}{Shumeet Baluja}, \bibinfo{person}{Rohan Seth}, \bibinfo{person}{Dharshi Sivakumar}, \bibinfo{person}{Yushi Jing}, \bibinfo{person}{Jay Yagnik}, \bibinfo{person}{Shankar Kumar}, \bibinfo{person}{Deepak Ravichandran}, {and} \bibinfo{person}{Mohamed Aly}.} \bibinfo{year}{2008}\natexlab{}.
\newblock \showarticletitle{Video suggestion and discovery for youtube: taking random walks through the view graph}. In \bibinfo{booktitle}{\emph{Proceedings of the 17th international conference on World Wide Web}}. \bibinfo{pages}{895--904}.
\newblock


\bibitem[Bommasani et~al\mbox{.}(2021)]%
        {bommasani2021opportunities}
\bibfield{author}{\bibinfo{person}{Rishi Bommasani}, \bibinfo{person}{Drew~A Hudson}, \bibinfo{person}{Ehsan Adeli}, \bibinfo{person}{Russ Altman}, \bibinfo{person}{Simran Arora}, \bibinfo{person}{Sydney von Arx}, \bibinfo{person}{Michael~S Bernstein}, \bibinfo{person}{Jeannette Bohg}, \bibinfo{person}{Antoine Bosselut}, \bibinfo{person}{Emma Brunskill}, {et~al\mbox{.}}} \bibinfo{year}{2021}\natexlab{}.
\newblock \showarticletitle{On the opportunities and risks of foundation models}.
\newblock \bibinfo{journal}{\emph{arXiv preprint arXiv:2108.07258}} (\bibinfo{year}{2021}).
\newblock


\bibitem[Cai et~al\mbox{.}(2025)]%
        {cai2025boosting}
\bibfield{author}{\bibinfo{person}{Rui Cai}, \bibinfo{person}{Chao Wang}, \bibinfo{person}{Qianyi Cai}, \bibinfo{person}{Dazhong Shen}, {and} \bibinfo{person}{Hui Xiong}.} \bibinfo{year}{2025}\natexlab{}.
\newblock \showarticletitle{Boosting Knowledge Graph-based Recommendations through Confidence-Aware Augmentation with Large Language Models}.
\newblock \bibinfo{journal}{\emph{arXiv preprint arXiv:2502.03715}} (\bibinfo{year}{2025}).
\newblock


\bibitem[Chen et~al\mbox{.}(2025a)]%
        {chen2025hierarchical}
\bibfield{author}{\bibinfo{person}{Jinpeng Chen}, \bibinfo{person}{Jianxiang He}, \bibinfo{person}{Huan Li}, \bibinfo{person}{Senzhang Wang}, \bibinfo{person}{Yuan Cao}, \bibinfo{person}{Kaimin Wei}, \bibinfo{person}{Zhenye Yang}, {and} \bibinfo{person}{Ye Ji}.} \bibinfo{year}{2025}\natexlab{a}.
\newblock \showarticletitle{Hierarchical Intent-guided Optimization with Pluggable LLM-Driven Semantics for Session-based Recommendation}. In \bibinfo{booktitle}{\emph{Proceedings of the 48th International ACM SIGIR Conference on Research and Development in Information Retrieval}}. \bibinfo{pages}{1655--1665}.
\newblock


\bibitem[Chen and Suzumura(2024)]%
        {chen2024prompting}
\bibfield{author}{\bibinfo{person}{Junyi Chen} {and} \bibinfo{person}{Toyotaro Suzumura}.} \bibinfo{year}{2024}\natexlab{}.
\newblock \showarticletitle{A Prompting-Based Representation Learning Method for Recommendation with Large Language Models}.
\newblock \bibinfo{journal}{\emph{arXiv preprint arXiv:2409.16674}} (\bibinfo{year}{2024}).
\newblock


\bibitem[Chen et~al\mbox{.}(2025b)]%
        {chen2025corona}
\bibfield{author}{\bibinfo{person}{Junze Chen}, \bibinfo{person}{Xinjie Yang}, \bibinfo{person}{Cheng Yang}, \bibinfo{person}{Junfei Bao}, \bibinfo{person}{Zeyuan Guo}, \bibinfo{person}{Yawen Li}, {and} \bibinfo{person}{Chuan Shi}.} \bibinfo{year}{2025}\natexlab{b}.
\newblock \showarticletitle{CORONA: A Coarse-to-Fine Framework for Graph-based Recommendation with Large Language Models}. In \bibinfo{booktitle}{\emph{Proceedings of the 48th International ACM SIGIR Conference on Research and Development in Information Retrieval}}. \bibinfo{pages}{2048--2058}.
\newblock


\bibitem[Chen et~al\mbox{.}(2024)]%
        {chen2024leverage}
\bibfield{author}{\bibinfo{person}{Yongming Chen}, \bibinfo{person}{Miner Chen}, \bibinfo{person}{Ye Zhu}, \bibinfo{person}{Juan Pei}, \bibinfo{person}{Siyu Chen}, \bibinfo{person}{Yu Zhou}, \bibinfo{person}{Yi Wang}, \bibinfo{person}{Yifan Zhou}, \bibinfo{person}{Hao Li}, {and} \bibinfo{person}{Songan Zhang}.} \bibinfo{year}{2024}\natexlab{}.
\newblock \showarticletitle{Leverage Knowledge Graph and Large Language Model for Law Article Recommendation: A Case Study of Chinese Criminal Law}.
\newblock \bibinfo{journal}{\emph{arXiv preprint arXiv:2410.04949}} (\bibinfo{year}{2024}).
\newblock


\bibitem[Covington et~al\mbox{.}(2016)]%
        {covington2016deep}
\bibfield{author}{\bibinfo{person}{Paul Covington}, \bibinfo{person}{Jay Adams}, {and} \bibinfo{person}{Emre Sargin}.} \bibinfo{year}{2016}\natexlab{}.
\newblock \showarticletitle{Deep neural networks for youtube recommendations}. In \bibinfo{booktitle}{\emph{Proceedings of the 10th ACM conference on recommender systems}}. \bibinfo{pages}{191--198}.
\newblock


\bibitem[Cui et~al\mbox{.}(2025)]%
        {cui2025comprehending}
\bibfield{author}{\bibinfo{person}{Ziqiang Cui}, \bibinfo{person}{Yunpeng Weng}, \bibinfo{person}{Xing Tang}, \bibinfo{person}{Fuyuan Lyu}, \bibinfo{person}{Dugang Liu}, \bibinfo{person}{Xiuqiang He}, {and} \bibinfo{person}{Chen Ma}.} \bibinfo{year}{2025}\natexlab{}.
\newblock \showarticletitle{Comprehending knowledge graphs with large language models for recommender systems}. In \bibinfo{booktitle}{\emph{Proceedings of the 48th International ACM SIGIR Conference on Research and Development in Information Retrieval}}. \bibinfo{pages}{1229--1239}.
\newblock


\bibitem[Damianou et~al\mbox{.}(2024)]%
        {damianou2024towards}
\bibfield{author}{\bibinfo{person}{Andreas Damianou}, \bibinfo{person}{Francesco Fabbri}, \bibinfo{person}{Paul Gigioli}, \bibinfo{person}{Marco De~Nadai}, \bibinfo{person}{Alice Wang}, \bibinfo{person}{Enrico Palumbo}, {and} \bibinfo{person}{Mounia Lalmas}.} \bibinfo{year}{2024}\natexlab{}.
\newblock \showarticletitle{Towards Graph Foundation Models for Personalization}. In \bibinfo{booktitle}{\emph{Companion Proceedings of the ACM on Web Conference 2024}}. \bibinfo{pages}{1798--1802}.
\newblock


\bibitem[Du et~al\mbox{.}(2024)]%
        {du2024large}
\bibfield{author}{\bibinfo{person}{Yingpeng Du}, \bibinfo{person}{Ziyan Wang}, \bibinfo{person}{Zhu Sun}, \bibinfo{person}{Haoyan Chua}, \bibinfo{person}{Hongzhi Liu}, \bibinfo{person}{Zhonghai Wu}, \bibinfo{person}{Yining Ma}, \bibinfo{person}{Jie Zhang}, {and} \bibinfo{person}{Youchen Sun}.} \bibinfo{year}{2024}\natexlab{}.
\newblock \showarticletitle{Large Language Model with Graph Convolution for Recommendation}.
\newblock \bibinfo{journal}{\emph{arXiv preprint arXiv:2402.08859}} (\bibinfo{year}{2024}).
\newblock


\bibitem[Ebrat et~al\mbox{.}(2025)]%
        {ebrat2025end}
\bibfield{author}{\bibinfo{person}{Danial Ebrat}, \bibinfo{person}{Tina Aminian}, \bibinfo{person}{Sepideh Ahmadian}, {and} \bibinfo{person}{Luis Rueda}.} \bibinfo{year}{2025}\natexlab{}.
\newblock \showarticletitle{End-to-End Personalization: Unifying Recommender Systems with Large Language Models}.
\newblock \bibinfo{journal}{\emph{arXiv preprint arXiv:2508.01514}} (\bibinfo{year}{2025}).
\newblock


\bibitem[Gao et~al\mbox{.}(2023)]%
        {gao2023survey}
\bibfield{author}{\bibinfo{person}{Chen Gao}, \bibinfo{person}{Yu Zheng}, \bibinfo{person}{Nian Li}, \bibinfo{person}{Yinfeng Li}, \bibinfo{person}{Yingrong Qin}, \bibinfo{person}{Jinghua Piao}, \bibinfo{person}{Yuhan Quan}, \bibinfo{person}{Jianxin Chang}, \bibinfo{person}{Depeng Jin}, \bibinfo{person}{Xiangnan He}, {et~al\mbox{.}}} \bibinfo{year}{2023}\natexlab{}.
\newblock \showarticletitle{A survey of graph neural networks for recommender systems: Challenges, methods, and directions}.
\newblock \bibinfo{journal}{\emph{ACM Transactions on Recommender Systems}} \bibinfo{volume}{1}, \bibinfo{number}{1} (\bibinfo{year}{2023}), \bibinfo{pages}{1--51}.
\newblock


\bibitem[Guan et~al\mbox{.}(2024)]%
        {guan2024enhancing}
\bibfield{author}{\bibinfo{person}{Zhong Guan}, \bibinfo{person}{Likang Wu}, \bibinfo{person}{Hongke Zhao}, \bibinfo{person}{Ming He}, {and} \bibinfo{person}{Jianpin Fan}.} \bibinfo{year}{2024}\natexlab{}.
\newblock \showarticletitle{Enhancing Collaborative Semantics of Language Model-Driven Recommendations via Graph-Aware Learning}.
\newblock \bibinfo{journal}{\emph{arXiv preprint arXiv:2406.13235}} (\bibinfo{year}{2024}).
\newblock


\bibitem[Guo et~al\mbox{.}(2017)]%
        {guo2017deepfm}
\bibfield{author}{\bibinfo{person}{Huifeng Guo}, \bibinfo{person}{Ruiming Tang}, \bibinfo{person}{Yunming Ye}, \bibinfo{person}{Zhenguo Li}, {and} \bibinfo{person}{Xiuqiang He}.} \bibinfo{year}{2017}\natexlab{}.
\newblock \showarticletitle{DeepFM: a factorization-machine based neural network for CTR prediction}.
\newblock \bibinfo{journal}{\emph{arXiv preprint arXiv:1703.04247}} (\bibinfo{year}{2017}).
\newblock


\bibitem[Guo et~al\mbox{.}(2024)]%
        {guo2024integrating}
\bibfield{author}{\bibinfo{person}{Naicheng Guo}, \bibinfo{person}{Hongwei Cheng}, \bibinfo{person}{Qianqiao Liang}, \bibinfo{person}{Linxun Chen}, {and} \bibinfo{person}{Bing Han}.} \bibinfo{year}{2024}\natexlab{}.
\newblock \showarticletitle{Integrating Large Language Models with Graphical Session-Based Recommendation}.
\newblock \bibinfo{journal}{\emph{arXiv preprint arXiv:2402.16539}} (\bibinfo{year}{2024}).
\newblock


\bibitem[Guo(2025)]%
        {pv3b-bx85-25}
\bibfield{author}{\bibinfo{person}{Xu Guo}.} \bibinfo{year}{2025}\natexlab{}.
\newblock \bibinfo{title}{HeLLM: Multi-Modal Hypergraph Enhanced LLM Learning for Recommendation}.
\newblock
\newblock


\bibitem[Hayati et~al\mbox{.}(2020)]%
        {hayati2020inspired}
\bibfield{author}{\bibinfo{person}{Shirley~Anugrah Hayati}, \bibinfo{person}{Dongyeop Kang}, \bibinfo{person}{Qingxiaoyang Zhu}, \bibinfo{person}{Weiyan Shi}, {and} \bibinfo{person}{Zhou Yu}.} \bibinfo{year}{2020}\natexlab{}.
\newblock \showarticletitle{Inspired: Toward sociable recommendation dialog systems}.
\newblock \bibinfo{journal}{\emph{arXiv preprint arXiv:2009.14306}} (\bibinfo{year}{2020}).
\newblock


\bibitem[He et~al\mbox{.}(2020)]%
        {he2020lightgcn}
\bibfield{author}{\bibinfo{person}{Xiangnan He}, \bibinfo{person}{Kuan Deng}, \bibinfo{person}{Xiang Wang}, \bibinfo{person}{Yan Li}, \bibinfo{person}{Yongdong Zhang}, {and} \bibinfo{person}{Meng Wang}.} \bibinfo{year}{2020}\natexlab{}.
\newblock \showarticletitle{Lightgcn: Simplifying and powering graph convolution network for recommendation}. In \bibinfo{booktitle}{\emph{Proceedings of the 43rd International ACM SIGIR conference on research and development in Information Retrieval}}. \bibinfo{pages}{639--648}.
\newblock


\bibitem[He et~al\mbox{.}(2017)]%
        {he2017neural}
\bibfield{author}{\bibinfo{person}{Xiangnan He}, \bibinfo{person}{Lizi Liao}, \bibinfo{person}{Hanwang Zhang}, \bibinfo{person}{Liqiang Nie}, \bibinfo{person}{Xia Hu}, {and} \bibinfo{person}{Tat-Seng Chua}.} \bibinfo{year}{2017}\natexlab{}.
\newblock \showarticletitle{Neural collaborative filtering}. In \bibinfo{booktitle}{\emph{Proceedings of the 26th international conference on world wide web}}. \bibinfo{pages}{173--182}.
\newblock


\bibitem[Hidasi et~al\mbox{.}(2015)]%
        {hidasi2015session}
\bibfield{author}{\bibinfo{person}{Bal{\'a}zs Hidasi}, \bibinfo{person}{Alexandros Karatzoglou}, \bibinfo{person}{Linas Baltrunas}, {and} \bibinfo{person}{Domonkos Tikk}.} \bibinfo{year}{2015}\natexlab{}.
\newblock \showarticletitle{Session-based recommendations with recurrent neural networks}.
\newblock \bibinfo{journal}{\emph{arXiv preprint arXiv:1511.06939}} (\bibinfo{year}{2015}).
\newblock


\bibitem[Hu et~al\mbox{.}(2008)]%
        {hu2008collaborative}
\bibfield{author}{\bibinfo{person}{Yifan Hu}, \bibinfo{person}{Yehuda Koren}, {and} \bibinfo{person}{Chris Volinsky}.} \bibinfo{year}{2008}\natexlab{}.
\newblock \showarticletitle{Collaborative filtering for implicit feedback datasets}. In \bibinfo{booktitle}{\emph{2008 Eighth IEEE international conference on data mining}}. Ieee, \bibinfo{pages}{263--272}.
\newblock


\bibitem[Hu et~al\mbox{.}(2025)]%
        {hu2025bridging}
\bibfield{author}{\bibinfo{person}{Zheng Hu}, \bibinfo{person}{Zhe Li}, \bibinfo{person}{Ziyun Jiao}, \bibinfo{person}{Satoshi Nakagawa}, \bibinfo{person}{Jiawen Deng}, \bibinfo{person}{Shimin Cai}, \bibinfo{person}{Tao Zhou}, {and} \bibinfo{person}{Fuji Ren}.} \bibinfo{year}{2025}\natexlab{}.
\newblock \showarticletitle{Bridging the user-side knowledge gap in knowledge-aware recommendations with large language models}. In \bibinfo{booktitle}{\emph{Proceedings of the AAAI Conference on Artificial Intelligence}}, Vol.~\bibinfo{volume}{39}. \bibinfo{pages}{11799--11807}.
\newblock


\bibitem[Huang et~al\mbox{.}(2025b)]%
        {huang2025survey}
\bibfield{author}{\bibinfo{person}{Chengkai Huang}, \bibinfo{person}{Hongtao Huang}, \bibinfo{person}{Tong Yu}, \bibinfo{person}{Kaige Xie}, \bibinfo{person}{Junda Wu}, \bibinfo{person}{Shuai Zhang}, \bibinfo{person}{Julian Mcauley}, \bibinfo{person}{Dietmar Jannach}, {and} \bibinfo{person}{Lina Yao}.} \bibinfo{year}{2025}\natexlab{b}.
\newblock \showarticletitle{A Survey of Foundation Model-Powered Recommender Systems: From Feature-Based, Generative to Agentic Paradigms}.
\newblock \bibinfo{journal}{\emph{arXiv preprint arXiv:2504.16420}} (\bibinfo{year}{2025}).
\newblock


\bibitem[Huang et~al\mbox{.}(2025a)]%
        {huang2025comprehensive}
\bibfield{author}{\bibinfo{person}{Junjie Huang}, \bibinfo{person}{Jizheng Chen}, \bibinfo{person}{Jianghao Lin}, \bibinfo{person}{Jiarui Qin}, \bibinfo{person}{Ziming Feng}, \bibinfo{person}{Weinan Zhang}, {and} \bibinfo{person}{Yong Yu}.} \bibinfo{year}{2025}\natexlab{a}.
\newblock \showarticletitle{A comprehensive survey on retrieval methods in recommender systems}.
\newblock \bibinfo{journal}{\emph{ACM Transactions on Information Systems}} \bibinfo{volume}{44}, \bibinfo{number}{1} (\bibinfo{year}{2025}), \bibinfo{pages}{1--43}.
\newblock


\bibitem[Isinkaye et~al\mbox{.}(2015)]%
        {isinkaye2015recommendation}
\bibfield{author}{\bibinfo{person}{FO Isinkaye}, \bibinfo{person}{YO Folajimi}, {and} \bibinfo{person}{BA Ojokoh}.} \bibinfo{year}{2015}\natexlab{}.
\newblock \showarticletitle{Recommendation systems: Principles, methods and evaluation}.
\newblock \bibinfo{journal}{\emph{Egyptian informatics journal}} \bibinfo{volume}{16}, \bibinfo{number}{3} (\bibinfo{year}{2015}), \bibinfo{pages}{261--273}.
\newblock


\bibitem[Jeon et~al\mbox{.}(2024)]%
        {jeon2024topic}
\bibfield{author}{\bibinfo{person}{Minhye Jeon}, \bibinfo{person}{Seokho Ahn}, {and} \bibinfo{person}{Young-Duk Seo}.} \bibinfo{year}{2024}\natexlab{}.
\newblock \showarticletitle{Topic-Aware Knowledge Graph with Large Language Models for Interoperability in Recommender Systems}.
\newblock \bibinfo{journal}{\emph{arXiv preprint arXiv:2412.20163}} (\bibinfo{year}{2024}).
\newblock


\bibitem[Jia et~al\mbox{.}(2025)]%
        {jia2025hetgcot}
\bibfield{author}{\bibinfo{person}{Runsong Jia}, \bibinfo{person}{Mengjia Wu}, \bibinfo{person}{Ying Ding}, \bibinfo{person}{Jie Lu}, {and} \bibinfo{person}{Yi Zhang}.} \bibinfo{year}{2025}\natexlab{}.
\newblock \showarticletitle{HetGCoT-Rec: Heterogeneous Graph-Enhanced Chain-of-Thought LLM Reasoning for Journal Recommendation}.
\newblock \bibinfo{journal}{\emph{arXiv preprint arXiv:2501.01203}} (\bibinfo{year}{2025}).
\newblock


\bibitem[Jin et~al\mbox{.}(2024)]%
        {jin2024large}
\bibfield{author}{\bibinfo{person}{Bowen Jin}, \bibinfo{person}{Gang Liu}, \bibinfo{person}{Chi Han}, \bibinfo{person}{Meng Jiang}, \bibinfo{person}{Heng Ji}, {and} \bibinfo{person}{Jiawei Han}.} \bibinfo{year}{2024}\natexlab{}.
\newblock \showarticletitle{Large language models on graphs: A comprehensive survey}.
\newblock \bibinfo{journal}{\emph{IEEE Transactions on Knowledge and Data Engineering}} (\bibinfo{year}{2024}).
\newblock


\bibitem[Joachims et~al\mbox{.}(2017)]%
        {joachims2017unbiased}
\bibfield{author}{\bibinfo{person}{Thorsten Joachims}, \bibinfo{person}{Adith Swaminathan}, {and} \bibinfo{person}{Tobias Schnabel}.} \bibinfo{year}{2017}\natexlab{}.
\newblock \showarticletitle{Unbiased learning-to-rank with biased feedback}. In \bibinfo{booktitle}{\emph{Proceedings of the tenth ACM international conference on web search and data mining}}. \bibinfo{pages}{781--789}.
\newblock


\bibitem[Kang and McAuley(2018)]%
        {kang2018self}
\bibfield{author}{\bibinfo{person}{Wang-Cheng Kang} {and} \bibinfo{person}{Julian McAuley}.} \bibinfo{year}{2018}\natexlab{}.
\newblock \showarticletitle{Self-attentive sequential recommendation}. In \bibinfo{booktitle}{\emph{2018 IEEE international conference on data mining (ICDM)}}. IEEE, \bibinfo{pages}{197--206}.
\newblock


\bibitem[Koren et~al\mbox{.}(2009)]%
        {koren2009matrix}
\bibfield{author}{\bibinfo{person}{Yehuda Koren}, \bibinfo{person}{Robert Bell}, {and} \bibinfo{person}{Chris Volinsky}.} \bibinfo{year}{2009}\natexlab{}.
\newblock \showarticletitle{Matrix factorization techniques for recommender systems}.
\newblock \bibinfo{journal}{\emph{Computer}} \bibinfo{volume}{42}, \bibinfo{number}{8} (\bibinfo{year}{2009}), \bibinfo{pages}{30--37}.
\newblock


\bibitem[Li et~al\mbox{.}(2025a)]%
        {li2025counterfactual}
\bibfield{author}{\bibinfo{person}{Guanrong Li}, \bibinfo{person}{Haolin Yang}, \bibinfo{person}{Xinyu Liu}, \bibinfo{person}{Zhen Wu}, {and} \bibinfo{person}{Xinyu Dai}.} \bibinfo{year}{2025}\natexlab{a}.
\newblock \showarticletitle{Counterfactual Language Reasoning for Explainable Recommendation Systems}.
\newblock \bibinfo{journal}{\emph{arXiv preprint arXiv:2503.08051}} (\bibinfo{year}{2025}).
\newblock


\bibitem[Li et~al\mbox{.}(2024)]%
        {li2024learning}
\bibfield{author}{\bibinfo{person}{Qingyao Li}, \bibinfo{person}{Wei Xia}, \bibinfo{person}{Kounianhua Du}, \bibinfo{person}{Qiji Zhang}, \bibinfo{person}{Weinan Zhang}, \bibinfo{person}{Ruiming Tang}, {and} \bibinfo{person}{Yong Yu}.} \bibinfo{year}{2024}\natexlab{}.
\newblock \showarticletitle{Learning Structure and Knowledge Aware Representation with Large Language Models for Concept Recommendation}.
\newblock \bibinfo{journal}{\emph{arXiv preprint arXiv:2405.12442}} (\bibinfo{year}{2024}).
\newblock


\bibitem[Li et~al\mbox{.}(2018)]%
        {li2018towards}
\bibfield{author}{\bibinfo{person}{Raymond Li}, \bibinfo{person}{Samira Ebrahimi~Kahou}, \bibinfo{person}{Hannes Schulz}, \bibinfo{person}{Vincent Michalski}, \bibinfo{person}{Laurent Charlin}, {and} \bibinfo{person}{Chris Pal}.} \bibinfo{year}{2018}\natexlab{}.
\newblock \showarticletitle{Towards deep conversational recommendations}.
\newblock \bibinfo{journal}{\emph{Advances in neural information processing systems}}  \bibinfo{volume}{31} (\bibinfo{year}{2018}).
\newblock


\bibitem[Li et~al\mbox{.}(2023)]%
        {li2023survey}
\bibfield{author}{\bibinfo{person}{Yuhan Li}, \bibinfo{person}{Zhixun Li}, \bibinfo{person}{Peisong Wang}, \bibinfo{person}{Jia Li}, \bibinfo{person}{Xiangguo Sun}, \bibinfo{person}{Hong Cheng}, {and} \bibinfo{person}{Jeffrey~Xu Yu}.} \bibinfo{year}{2023}\natexlab{}.
\newblock \showarticletitle{A survey of graph meets large language model: Progress and future directions}.
\newblock \bibinfo{journal}{\emph{arXiv preprint arXiv:2311.12399}} (\bibinfo{year}{2023}).
\newblock


\bibitem[Li et~al\mbox{.}(2025b)]%
        {li2025g}
\bibfield{author}{\bibinfo{person}{Yuhan Li}, \bibinfo{person}{Xinni Zhang}, \bibinfo{person}{Linhao Luo}, \bibinfo{person}{Heng Chang}, \bibinfo{person}{Yuxiang Ren}, \bibinfo{person}{Irwin King}, {and} \bibinfo{person}{Jia Li}.} \bibinfo{year}{2025}\natexlab{b}.
\newblock \showarticletitle{G-refer: Graph retrieval-augmented large language model for explainable recommendation}. In \bibinfo{booktitle}{\emph{Proceedings of the ACM on Web Conference 2025}}. \bibinfo{pages}{240--251}.
\newblock


\bibitem[Lin et~al\mbox{.}(2025)]%
        {lin2025can}
\bibfield{author}{\bibinfo{person}{Jianghao Lin}, \bibinfo{person}{Xinyi Dai}, \bibinfo{person}{Yunjia Xi}, \bibinfo{person}{Weiwen Liu}, \bibinfo{person}{Bo Chen}, \bibinfo{person}{Hao Zhang}, \bibinfo{person}{Yong Liu}, \bibinfo{person}{Chuhan Wu}, \bibinfo{person}{Xiangyang Li}, \bibinfo{person}{Chenxu Zhu}, {et~al\mbox{.}}} \bibinfo{year}{2025}\natexlab{}.
\newblock \showarticletitle{How can recommender systems benefit from large language models: A survey}.
\newblock \bibinfo{journal}{\emph{ACM Transactions on Information Systems}} \bibinfo{volume}{43}, \bibinfo{number}{2} (\bibinfo{year}{2025}), \bibinfo{pages}{1--47}.
\newblock


\bibitem[Liu et~al\mbox{.}(2023a)]%
        {liu2023understanding}
\bibfield{author}{\bibinfo{person}{Fan Liu}, \bibinfo{person}{Yaqi Liu}, \bibinfo{person}{Huilin Chen}, \bibinfo{person}{Zhiyong Cheng}, \bibinfo{person}{Liqiang Nie}, {and} \bibinfo{person}{Mohan Kankanhalli}.} \bibinfo{year}{2023}\natexlab{a}.
\newblock \showarticletitle{Understanding Before Recommendation: Semantic Aspect-Aware Review Exploitation via Large Language Models}.
\newblock \bibinfo{journal}{\emph{ACM Transactions on Information Systems}} (\bibinfo{year}{2023}).
\newblock


\bibitem[Liu et~al\mbox{.}(2023b)]%
        {liu2023towards}
\bibfield{author}{\bibinfo{person}{Jiawei Liu}, \bibinfo{person}{Cheng Yang}, \bibinfo{person}{Zhiyuan Lu}, \bibinfo{person}{Junze Chen}, \bibinfo{person}{Yibo Li}, \bibinfo{person}{Mengmei Zhang}, \bibinfo{person}{Ting Bai}, \bibinfo{person}{Yuan Fang}, \bibinfo{person}{Lichao Sun}, \bibinfo{person}{Philip~S Yu}, {et~al\mbox{.}}} \bibinfo{year}{2023}\natexlab{b}.
\newblock \showarticletitle{Towards graph foundation models: A survey and beyond}.
\newblock \bibinfo{journal}{\emph{arXiv preprint arXiv:2310.11829}} (\bibinfo{year}{2023}).
\newblock


\bibitem[Luo et~al\mbox{.}(2025)]%
        {luo2025llm}
\bibfield{author}{\bibinfo{person}{Hailong Luo}, \bibinfo{person}{Bin Wu}, \bibinfo{person}{Hongyong Jia}, \bibinfo{person}{Qingqing Zhu}, {and} \bibinfo{person}{Lianlei Shan}.} \bibinfo{year}{2025}\natexlab{}.
\newblock \showarticletitle{Llm-cot enhanced graph neural recommendation with harmonized group policy optimization}.
\newblock \bibinfo{journal}{\emph{arXiv preprint arXiv:2505.12396}} (\bibinfo{year}{2025}).
\newblock


\bibitem[Ma et~al\mbox{.}(2024a)]%
        {ma2024triple}
\bibfield{author}{\bibinfo{person}{Luyi Ma}, \bibinfo{person}{Xiaohan Li}, \bibinfo{person}{Zezhong Fan}, \bibinfo{person}{Jianpeng Xu}, \bibinfo{person}{Jason Cho}, \bibinfo{person}{Praveen Kanumala}, \bibinfo{person}{Kaushiki Nag}, \bibinfo{person}{Sushant Kumar}, {and} \bibinfo{person}{Kannan Achan}.} \bibinfo{year}{2024}\natexlab{a}.
\newblock \showarticletitle{Triple Modality Fusion: Aligning Visual, Textual, and Graph Data with Large Language Models for Multi-Behavior Recommendations}.
\newblock \bibinfo{journal}{\emph{arXiv preprint arXiv:2410.12228}} (\bibinfo{year}{2024}).
\newblock


\bibitem[Ma et~al\mbox{.}(2024b)]%
        {ma2024xrec}
\bibfield{author}{\bibinfo{person}{Qiyao Ma}, \bibinfo{person}{Xubin Ren}, {and} \bibinfo{person}{Chao Huang}.} \bibinfo{year}{2024}\natexlab{b}.
\newblock \showarticletitle{XRec: Large Language Models for Explainable Recommendation}.
\newblock \bibinfo{journal}{\emph{arXiv preprint arXiv:2406.02377}} (\bibinfo{year}{2024}).
\newblock


\bibitem[Mao et~al\mbox{.}(2024a)]%
        {mao2024graph}
\bibfield{author}{\bibinfo{person}{Haitao Mao}, \bibinfo{person}{Zhikai Chen}, \bibinfo{person}{Wenzhuo Tang}, \bibinfo{person}{Jianan Zhao}, \bibinfo{person}{Yao Ma}, \bibinfo{person}{Tong Zhao}, \bibinfo{person}{Neil Shah}, \bibinfo{person}{Mikhail Galkin}, {and} \bibinfo{person}{Jiliang Tang}.} \bibinfo{year}{2024}\natexlab{a}.
\newblock \showarticletitle{Graph foundation models}.
\newblock \bibinfo{journal}{\emph{CoRR}} (\bibinfo{year}{2024}).
\newblock


\bibitem[Mao et~al\mbox{.}(2024b)]%
        {mao2024position}
\bibfield{author}{\bibinfo{person}{Haitao Mao}, \bibinfo{person}{Zhikai Chen}, \bibinfo{person}{Wenzhuo Tang}, \bibinfo{person}{Jianan Zhao}, \bibinfo{person}{Yao Ma}, \bibinfo{person}{Tong Zhao}, \bibinfo{person}{Neil Shah}, \bibinfo{person}{Mikhail Galkin}, {and} \bibinfo{person}{Jiliang Tang}.} \bibinfo{year}{2024}\natexlab{b}.
\newblock \showarticletitle{Position: Graph foundation models are already here}. In \bibinfo{booktitle}{\emph{Forty-first International Conference on Machine Learning}}.
\newblock


\bibitem[Mei and Zhang(2023)]%
        {mei2023lightlm}
\bibfield{author}{\bibinfo{person}{Kai Mei} {and} \bibinfo{person}{Yongfeng Zhang}.} \bibinfo{year}{2023}\natexlab{}.
\newblock \showarticletitle{LightLM: a lightweight deep and narrow language model for generative recommendation}.
\newblock \bibinfo{journal}{\emph{arXiv preprint arXiv:2310.17488}} (\bibinfo{year}{2023}).
\newblock


\bibitem[Meng et~al\mbox{.}(2025a)]%
        {Meng_Meng_Jin_Lin_Wu_2025}
\bibfield{author}{\bibinfo{person}{Fanshen Meng}, \bibinfo{person}{Zhenhua Meng}, \bibinfo{person}{Ru Jin}, \bibinfo{person}{Rongheng Lin}, {and} \bibinfo{person}{Budan Wu}.} \bibinfo{year}{2025}\natexlab{a}.
\newblock \showarticletitle{DOGE: LLMs-Enhanced Hyper-Knowledge Graph Recommender for Multimodal Recommendation}.
\newblock \bibinfo{journal}{\emph{Proceedings of the AAAI Conference on Artificial Intelligence}} \bibinfo{volume}{39}, \bibinfo{number}{12} (\bibinfo{date}{Apr.} \bibinfo{year}{2025}), \bibinfo{pages}{12399--12407}.
\newblock


\bibitem[Meng et~al\mbox{.}(2025b)]%
        {meng2025kerag_r}
\bibfield{author}{\bibinfo{person}{Zeyuan Meng}, \bibinfo{person}{Zixuan Yi}, {and} \bibinfo{person}{Iadh Ounis}.} \bibinfo{year}{2025}\natexlab{b}.
\newblock \showarticletitle{KERAG\_R: Knowledge-Enhanced Retrieval-Augmented Generation for Recommendation}.
\newblock \bibinfo{journal}{\emph{arXiv preprint arXiv:2507.05863}} (\bibinfo{year}{2025}).
\newblock


\bibitem[Mohbat and Zaki(2025)]%
        {mohbat2025kerl}
\bibfield{author}{\bibinfo{person}{Fnu Mohbat} {and} \bibinfo{person}{Mohammed~J Zaki}.} \bibinfo{year}{2025}\natexlab{}.
\newblock \showarticletitle{{KERL}: Knowledge-Enhanced Personalized Recipe Recommendation using Large Language Models}. In \bibinfo{booktitle}{\emph{Proceedings of the 63rd Annual Meeting of the Association for Computational Linguistics (Volume 1: Long Papers)}}, \bibfield{editor}{\bibinfo{person}{Wanxiang Che}, \bibinfo{person}{Joyce Nabende}, \bibinfo{person}{Ekaterina Shutova}, {and} \bibinfo{person}{Mohammad~Taher Pilehvar}} (Eds.). \bibinfo{publisher}{Association for Computational Linguistics}, \bibinfo{address}{Vienna, Austria}, \bibinfo{pages}{19125--19141}.
\newblock
\showISBNx{979-8-89176-251-0}


\bibitem[Nguyen et~al\mbox{.}(2025)]%
        {nguyen2025votegcl}
\bibfield{author}{\bibinfo{person}{Minh-Anh Nguyen}, \bibinfo{person}{Bao Nguyen}, \bibinfo{person}{Tuan~Anh Hoang}, \bibinfo{person}{Duc-Trong Le}, \bibinfo{person}{Dung~D Le}, {et~al\mbox{.}}} \bibinfo{year}{2025}\natexlab{}.
\newblock \showarticletitle{VoteGCL: Enhancing Graph-based Recommendations with Majority-Voting LLM-Rerank Augmentation}.
\newblock \bibinfo{journal}{\emph{arXiv preprint arXiv:2507.21563}} (\bibinfo{year}{2025}).
\newblock


\bibitem[Ning et~al\mbox{.}(2025)]%
        {ning2025retrieval}
\bibfield{author}{\bibinfo{person}{Liangbo Ning}, \bibinfo{person}{Wenqi Fan}, {and} \bibinfo{person}{Qing Li}.} \bibinfo{year}{2025}\natexlab{}.
\newblock \showarticletitle{Retrieval-Augmented Purifier for Robust LLM-Empowered Recommendation}.
\newblock \bibinfo{journal}{\emph{arXiv preprint arXiv:2504.02458}} (\bibinfo{year}{2025}).
\newblock


\bibitem[Peng et~al\mbox{.}(2024)]%
        {peng2024denoising}
\bibfield{author}{\bibinfo{person}{Yingtao Peng}, \bibinfo{person}{Chen Gao}, \bibinfo{person}{Yu Zhang}, \bibinfo{person}{Tangpeng Dan}, \bibinfo{person}{Xiaoyi Du}, \bibinfo{person}{Hengliang Luo}, \bibinfo{person}{Yong Li}, {and} \bibinfo{person}{Xiaofeng Meng}.} \bibinfo{year}{2024}\natexlab{}.
\newblock \showarticletitle{Denoising Alignment with Large Language Model for Recommendation}.
\newblock \bibinfo{journal}{\emph{ACM Transactions on Information Systems}} (\bibinfo{year}{2024}).
\newblock


\bibitem[Qiu et~al\mbox{.}(2024)]%
        {qiu2024unveiling}
\bibfield{author}{\bibinfo{person}{Zhangchi Qiu}, \bibinfo{person}{Linhao Luo}, \bibinfo{person}{Shirui Pan}, {and} \bibinfo{person}{Alan Wee-Chung Liew}.} \bibinfo{year}{2024}\natexlab{}.
\newblock \showarticletitle{Unveiling User Preferences: A Knowledge Graph and LLM-Driven Approach for Conversational Recommendation}.
\newblock \bibinfo{journal}{\emph{arXiv preprint arXiv:2411.14459}} (\bibinfo{year}{2024}).
\newblock


\bibitem[Qiu et~al\mbox{.}(2025)]%
        {qiu2025graph}
\bibfield{author}{\bibinfo{person}{Zhangchi Qiu}, \bibinfo{person}{Linhao Luo}, \bibinfo{person}{Zicheng Zhao}, \bibinfo{person}{Shirui Pan}, {and} \bibinfo{person}{Alan Wee-Chung Liew}.} \bibinfo{year}{2025}\natexlab{}.
\newblock \showarticletitle{Graph Retrieval-Augmented LLM for Conversational Recommendation Systems}. In \bibinfo{booktitle}{\emph{Pacific-Asia Conference on Knowledge Discovery and Data Mining}}. Springer, \bibinfo{pages}{344--355}.
\newblock


\bibitem[Ren et~al\mbox{.}(2024a)]%
        {ren2024survey}
\bibfield{author}{\bibinfo{person}{Xubin Ren}, \bibinfo{person}{Jiabin Tang}, \bibinfo{person}{Dawei Yin}, \bibinfo{person}{Nitesh Chawla}, {and} \bibinfo{person}{Chao Huang}.} \bibinfo{year}{2024}\natexlab{a}.
\newblock \showarticletitle{A survey of large language models for graphs}. In \bibinfo{booktitle}{\emph{Proceedings of the 30th ACM SIGKDD Conference on Knowledge Discovery and Data Mining}}. \bibinfo{pages}{6616--6626}.
\newblock


\bibitem[Ren et~al\mbox{.}(2024b)]%
        {ren2024representation}
\bibfield{author}{\bibinfo{person}{Xubin Ren}, \bibinfo{person}{Wei Wei}, \bibinfo{person}{Lianghao Xia}, \bibinfo{person}{Lixin Su}, \bibinfo{person}{Suqi Cheng}, \bibinfo{person}{Junfeng Wang}, \bibinfo{person}{Dawei Yin}, {and} \bibinfo{person}{Chao Huang}.} \bibinfo{year}{2024}\natexlab{b}.
\newblock \showarticletitle{Representation learning with large language models for recommendation}. In \bibinfo{booktitle}{\emph{Proceedings of the ACM on Web Conference 2024}}. \bibinfo{pages}{3464--3475}.
\newblock


\bibitem[Rendle et~al\mbox{.}(2012)]%
        {rendle2012bpr}
\bibfield{author}{\bibinfo{person}{Steffen Rendle}, \bibinfo{person}{Christoph Freudenthaler}, \bibinfo{person}{Zeno Gantner}, {and} \bibinfo{person}{Lars Schmidt-Thieme}.} \bibinfo{year}{2012}\natexlab{}.
\newblock \showarticletitle{BPR: Bayesian personalized ranking from implicit feedback}.
\newblock \bibinfo{journal}{\emph{arXiv preprint arXiv:1205.2618}} (\bibinfo{year}{2012}).
\newblock


\bibitem[Runfeng et~al\mbox{.}(2023)]%
        {runfeng2023lkpnr}
\bibfield{author}{\bibinfo{person}{Xie Runfeng}, \bibinfo{person}{Cui Xiangyang}, \bibinfo{person}{Yan Zhou}, \bibinfo{person}{Wang Xin}, \bibinfo{person}{Xuan Zhanwei}, \bibinfo{person}{Zhang Kai}, {et~al\mbox{.}}} \bibinfo{year}{2023}\natexlab{}.
\newblock \showarticletitle{Lkpnr: Llm and kg for personalized news recommendation framework}.
\newblock \bibinfo{journal}{\emph{arXiv preprint arXiv:2308.12028}} (\bibinfo{year}{2023}).
\newblock


\bibitem[Sakurai et~al\mbox{.}(2024)]%
        {sakurai2024llm}
\bibfield{author}{\bibinfo{person}{Keigo Sakurai}, \bibinfo{person}{Ren Togo}, \bibinfo{person}{Takahiro Ogawa}, {and} \bibinfo{person}{Miki Haseyama}.} \bibinfo{year}{2024}\natexlab{}.
\newblock \showarticletitle{LLM is Knowledge Graph Reasoner: LLM's Intuition-aware Knowledge Graph Reasoning for Cold-start Sequential Recommendation}.
\newblock \bibinfo{journal}{\emph{arXiv preprint arXiv:2412.12464}} (\bibinfo{year}{2024}).
\newblock


\bibitem[Schnabel et~al\mbox{.}(2016)]%
        {schnabel2016recommendations}
\bibfield{author}{\bibinfo{person}{Tobias Schnabel}, \bibinfo{person}{Adith Swaminathan}, \bibinfo{person}{Ashudeep Singh}, \bibinfo{person}{Navin Chandak}, {and} \bibinfo{person}{Thorsten Joachims}.} \bibinfo{year}{2016}\natexlab{}.
\newblock \showarticletitle{Recommendations as treatments: Debiasing learning and evaluation}. In \bibinfo{booktitle}{\emph{international conference on machine learning}}. PMLR, \bibinfo{pages}{1670--1679}.
\newblock


\bibitem[Shan et~al\mbox{.}(2024)]%
        {shan2024automatic}
\bibfield{author}{\bibinfo{person}{Rong Shan}, \bibinfo{person}{Jianghao Lin}, \bibinfo{person}{Chenxu Zhu}, \bibinfo{person}{Bo Chen}, \bibinfo{person}{Menghui Zhu}, \bibinfo{person}{Kangning Zhang}, \bibinfo{person}{Jieming Zhu}, \bibinfo{person}{Ruiming Tang}, \bibinfo{person}{Yong Yu}, {and} \bibinfo{person}{Weinan Zhang}.} \bibinfo{year}{2024}\natexlab{}.
\newblock \showarticletitle{An Automatic Graph Construction Framework based on Large Language Models for Recommendation}.
\newblock \bibinfo{journal}{\emph{arXiv preprint arXiv:2412.18241}} (\bibinfo{year}{2024}).
\newblock


\bibitem[Shen et~al\mbox{.}(2024)]%
        {shen2024exploring}
\bibfield{author}{\bibinfo{person}{Tingjia Shen}, \bibinfo{person}{Hao Wang}, \bibinfo{person}{Jiaqing Zhang}, \bibinfo{person}{Sirui Zhao}, \bibinfo{person}{Liangyue Li}, \bibinfo{person}{Zulong Chen}, \bibinfo{person}{Defu Lian}, {and} \bibinfo{person}{Enhong Chen}.} \bibinfo{year}{2024}\natexlab{}.
\newblock \showarticletitle{Exploring User Retrieval Integration towards Large Language Models for Cross-Domain Sequential Recommendation}.
\newblock \bibinfo{journal}{\emph{arXiv preprint arXiv:2406.03085}} (\bibinfo{year}{2024}).
\newblock


\bibitem[Shi et~al\mbox{.}(2020)]%
        {shi2020heterogeneous}
\bibfield{author}{\bibinfo{person}{Jinghan Shi}, \bibinfo{person}{Houye Ji}, \bibinfo{person}{Chuan Shi}, \bibinfo{person}{Xiao Wang}, \bibinfo{person}{Zhiqiang Zhang}, {and} \bibinfo{person}{Jun Zhou}.} \bibinfo{year}{2020}\natexlab{}.
\newblock \showarticletitle{Heterogeneous graph neural network for recommendation}.
\newblock \bibinfo{journal}{\emph{arXiv preprint arXiv:2009.00799}} (\bibinfo{year}{2020}).
\newblock


\bibitem[Shi et~al\mbox{.}(2025)]%
        {shi2025benefit}
\bibfield{author}{\bibinfo{person}{Teng Shi}, \bibinfo{person}{Weijie Yu}, \bibinfo{person}{Xiao Zhang}, \bibinfo{person}{Ming He}, \bibinfo{person}{Jianping Fan}, {and} \bibinfo{person}{Jun Xu}.} \bibinfo{year}{2025}\natexlab{}.
\newblock \showarticletitle{Benefit from Rich: Tackling Search Interaction Sparsity in Search Enhanced Recommendation}. In \bibinfo{booktitle}{\emph{Proceedings of the 34th ACM International Conference on Information and Knowledge Management}}. \bibinfo{pages}{2685--2695}.
\newblock


\bibitem[Su and Khoshgoftaar(2009)]%
        {su2009survey}
\bibfield{author}{\bibinfo{person}{Xiaoyuan Su} {and} \bibinfo{person}{Taghi~M Khoshgoftaar}.} \bibinfo{year}{2009}\natexlab{}.
\newblock \showarticletitle{A survey of collaborative filtering techniques}.
\newblock \bibinfo{journal}{\emph{Advances in artificial intelligence}} \bibinfo{volume}{2009}, \bibinfo{number}{1} (\bibinfo{year}{2009}), \bibinfo{pages}{421425}.
\newblock


\bibitem[Sun et~al\mbox{.}(2019)]%
        {sun2019bert4rec}
\bibfield{author}{\bibinfo{person}{Fei Sun}, \bibinfo{person}{Jun Liu}, \bibinfo{person}{Jian Wu}, \bibinfo{person}{Changhua Pei}, \bibinfo{person}{Xiao Lin}, \bibinfo{person}{Wenwu Ou}, {and} \bibinfo{person}{Peng Jiang}.} \bibinfo{year}{2019}\natexlab{}.
\newblock \showarticletitle{BERT4Rec: Sequential recommendation with bidirectional encoder representations from transformer}. In \bibinfo{booktitle}{\emph{Proceedings of the 28th ACM international conference on information and knowledge management}}. \bibinfo{pages}{1441--1450}.
\newblock


\bibitem[Tao et~al\mbox{.}(2020)]%
        {tao2020mgat}
\bibfield{author}{\bibinfo{person}{Zhulin Tao}, \bibinfo{person}{Yinwei Wei}, \bibinfo{person}{Xiang Wang}, \bibinfo{person}{Xiangnan He}, \bibinfo{person}{Xianglin Huang}, {and} \bibinfo{person}{Tat-Seng Chua}.} \bibinfo{year}{2020}\natexlab{}.
\newblock \showarticletitle{Mgat: Multimodal graph attention network for recommendation}.
\newblock \bibinfo{journal}{\emph{Information Processing \& Management}} \bibinfo{volume}{57}, \bibinfo{number}{5} (\bibinfo{year}{2020}), \bibinfo{pages}{102277}.
\newblock


\bibitem[Van~Meteren and Van~Someren(2000)]%
        {van2000using}
\bibfield{author}{\bibinfo{person}{Robin Van~Meteren} {and} \bibinfo{person}{Maarten Van~Someren}.} \bibinfo{year}{2000}\natexlab{}.
\newblock \showarticletitle{Using content-based filtering for recommendation}. In \bibinfo{booktitle}{\emph{Proceedings of the machine learning in the new information age: MLnet/ECML2000 workshop}}, Vol.~\bibinfo{volume}{30}. Barcelona, \bibinfo{pages}{47--56}.
\newblock


\bibitem[Wang et~al\mbox{.}(2018)]%
        {wang2018ripplenet}
\bibfield{author}{\bibinfo{person}{Hongwei Wang}, \bibinfo{person}{Fuzheng Zhang}, \bibinfo{person}{Jialin Wang}, \bibinfo{person}{Miao Zhao}, \bibinfo{person}{Wenjie Li}, \bibinfo{person}{Xing Xie}, {and} \bibinfo{person}{Minyi Guo}.} \bibinfo{year}{2018}\natexlab{}.
\newblock \showarticletitle{Ripplenet: Propagating user preferences on the knowledge graph for recommender systems}. In \bibinfo{booktitle}{\emph{Proceedings of the 27th ACM international conference on information and knowledge management}}. \bibinfo{pages}{417--426}.
\newblock


\bibitem[Wang et~al\mbox{.}(2024c)]%
        {10.1145/3640457.3688161}
\bibfield{author}{\bibinfo{person}{Jianling Wang}, \bibinfo{person}{Haokai Lu}, \bibinfo{person}{Yifan Liu}, \bibinfo{person}{He Ma}, \bibinfo{person}{Yueqi Wang}, \bibinfo{person}{Yang Gu}, \bibinfo{person}{Shuzhou Zhang}, \bibinfo{person}{Ningren Han}, \bibinfo{person}{Shuchao Bi}, \bibinfo{person}{Lexi Baugher}, \bibinfo{person}{Ed~H. Chi}, {and} \bibinfo{person}{Minmin Chen}.} \bibinfo{year}{2024}\natexlab{c}.
\newblock \showarticletitle{LLMs for User Interest Exploration in Large-scale Recommendation Systems}. In \bibinfo{booktitle}{\emph{Proceedings of the 18th ACM Conference on Recommender Systems}} (Bari, Italy) \emph{(\bibinfo{series}{RecSys '24})}. \bibinfo{publisher}{Association for Computing Machinery}, \bibinfo{address}{New York, NY, USA}, \bibinfo{pages}{872–877}.
\newblock
\showISBNx{9798400705052}


\bibitem[Wang et~al\mbox{.}(2025d)]%
        {wang2025towards}
\bibfield{author}{\bibinfo{person}{Leyao Wang}, \bibinfo{person}{Xutao Mao}, \bibinfo{person}{Xuhui Zhan}, \bibinfo{person}{Yuying Zhao}, \bibinfo{person}{Bo Ni}, \bibinfo{person}{Ryan~A Rossi}, \bibinfo{person}{Nesreen~K Ahmed}, {and} \bibinfo{person}{Tyler Derr}.} \bibinfo{year}{2025}\natexlab{d}.
\newblock \showarticletitle{Towards Bridging Review Sparsity in Recommendation with Textual Edge Graph Representation}.
\newblock \bibinfo{journal}{\emph{arXiv preprint arXiv:2508.01128}} (\bibinfo{year}{2025}).
\newblock


\bibitem[Wang et~al\mbox{.}(2024b)]%
        {wang2024enabling}
\bibfield{author}{\bibinfo{person}{Menghan Wang}, \bibinfo{person}{Yuchen Guo}, \bibinfo{person}{Duanfeng Zhang}, \bibinfo{person}{Jianian Jin}, \bibinfo{person}{Minnie Li}, \bibinfo{person}{Dan Schonfeld}, {and} \bibinfo{person}{Shawn Zhou}.} \bibinfo{year}{2024}\natexlab{b}.
\newblock \showarticletitle{Enabling Explainable Recommendation in E-commerce with LLM-powered Product Knowledge Graph}.
\newblock \bibinfo{journal}{\emph{arXiv preprint arXiv:2412.01837}} (\bibinfo{year}{2024}).
\newblock


\bibitem[Wang et~al\mbox{.}(2025b)]%
        {wang2025beyond}
\bibfield{author}{\bibinfo{person}{Minhao Wang}, \bibinfo{person}{Yunhang He}, \bibinfo{person}{Cong Xu}, \bibinfo{person}{Zhangchi Zhu}, {and} \bibinfo{person}{Wei Zhang}.} \bibinfo{year}{2025}\natexlab{b}.
\newblock \showarticletitle{Beyond Semantic Understanding: Preserving Collaborative Frequency Components in LLM-based Recommendation}.
\newblock \bibinfo{journal}{\emph{arXiv preprint arXiv:2508.10312}} (\bibinfo{year}{2025}).
\newblock


\bibitem[Wang et~al\mbox{.}(2025a)]%
        {wang-etal-2025-knowledge-graph}
\bibfield{author}{\bibinfo{person}{Shijie Wang}, \bibinfo{person}{Wenqi Fan}, \bibinfo{person}{Yue Feng}, \bibinfo{person}{Lin Shanru}, \bibinfo{person}{Xinyu Ma}, \bibinfo{person}{Shuaiqiang Wang}, {and} \bibinfo{person}{Dawei Yin}.} \bibinfo{year}{2025}\natexlab{a}.
\newblock \showarticletitle{Knowledge Graph Retrieval-Augmented Generation for {LLM}-based Recommendation}. In \bibinfo{booktitle}{\emph{Proceedings of the 63rd Annual Meeting of the Association for Computational Linguistics (Volume 1: Long Papers)}}, \bibfield{editor}{\bibinfo{person}{Wanxiang Che}, \bibinfo{person}{Joyce Nabende}, \bibinfo{person}{Ekaterina Shutova}, {and} \bibinfo{person}{Mohammad~Taher Pilehvar}} (Eds.). \bibinfo{publisher}{Association for Computational Linguistics}, \bibinfo{address}{Vienna, Austria}, \bibinfo{pages}{27152--27168}.
\newblock
\showISBNx{979-8-89176-251-0}


\bibitem[Wang et~al\mbox{.}(2025e)]%
        {wang2025unleashing}
\bibfield{author}{\bibinfo{person}{Shuyao Wang}, \bibinfo{person}{Zhi Zheng}, \bibinfo{person}{Yongduo Sui}, {and} \bibinfo{person}{Hui Xiong}.} \bibinfo{year}{2025}\natexlab{e}.
\newblock \showarticletitle{Unleashing the Power of Large Language Model for Denoising Recommendation}. In \bibinfo{booktitle}{\emph{Proceedings of the ACM on Web Conference 2025}}. \bibinfo{pages}{252--263}.
\newblock


\bibitem[Wang et~al\mbox{.}(2024a)]%
        {wang2024enhancing}
\bibfield{author}{\bibinfo{person}{Xinfeng Wang}, \bibinfo{person}{Jin Cui}, \bibinfo{person}{Fumiyo Fukumoto}, {and} \bibinfo{person}{Yoshimi Suzuki}.} \bibinfo{year}{2024}\natexlab{a}.
\newblock \showarticletitle{Enhancing High-order Interaction Awareness in LLM-based Recommender Model}. In \bibinfo{booktitle}{\emph{Proceedings of the 2024 Conference on Empirical Methods in Natural Language Processing}}. \bibinfo{pages}{11696--11711}.
\newblock


\bibitem[Wang et~al\mbox{.}(2019a)]%
        {wang2019kgat}
\bibfield{author}{\bibinfo{person}{Xiang Wang}, \bibinfo{person}{Xiangnan He}, \bibinfo{person}{Yixin Cao}, \bibinfo{person}{Meng Liu}, {and} \bibinfo{person}{Tat-Seng Chua}.} \bibinfo{year}{2019}\natexlab{a}.
\newblock \showarticletitle{Kgat: Knowledge graph attention network for recommendation}. In \bibinfo{booktitle}{\emph{Proceedings of the 25th ACM SIGKDD international conference on knowledge discovery \& data mining}}. \bibinfo{pages}{950--958}.
\newblock


\bibitem[Wang et~al\mbox{.}(2019b)]%
        {wang2019neural}
\bibfield{author}{\bibinfo{person}{Xiang Wang}, \bibinfo{person}{Xiangnan He}, \bibinfo{person}{Meng Wang}, \bibinfo{person}{Fuli Feng}, {and} \bibinfo{person}{Tat-Seng Chua}.} \bibinfo{year}{2019}\natexlab{b}.
\newblock \showarticletitle{Neural graph collaborative filtering}. In \bibinfo{booktitle}{\emph{Proceedings of the 42nd international ACM SIGIR conference on Research and development in Information Retrieval}}. \bibinfo{pages}{165--174}.
\newblock


\bibitem[Wang et~al\mbox{.}(2024d)]%
        {wang2024llm}
\bibfield{author}{\bibinfo{person}{Xinyuan Wang}, \bibinfo{person}{Liang Wu}, \bibinfo{person}{Liangjie Hong}, \bibinfo{person}{Hao Liu}, {and} \bibinfo{person}{Yanjie Fu}.} \bibinfo{year}{2024}\natexlab{d}.
\newblock \showarticletitle{LLM-Enhanced User-Item Interactions: Leveraging Edge Information for Optimized Recommendations}.
\newblock \bibinfo{journal}{\emph{arXiv preprint arXiv:2402.09617}} (\bibinfo{year}{2024}).
\newblock


\bibitem[Wang et~al\mbox{.}(2023)]%
        {wang2023enhancing}
\bibfield{author}{\bibinfo{person}{Yan Wang}, \bibinfo{person}{Zhixuan Chu}, \bibinfo{person}{Xin Ouyang}, \bibinfo{person}{Simeng Wang}, \bibinfo{person}{Hongyan Hao}, \bibinfo{person}{Yue Shen}, \bibinfo{person}{Jinjie Gu}, \bibinfo{person}{Siqiao Xue}, \bibinfo{person}{James~Y Zhang}, \bibinfo{person}{Qing Cui}, {et~al\mbox{.}}} \bibinfo{year}{2023}\natexlab{}.
\newblock \showarticletitle{Enhancing recommender systems with large language model reasoning graphs}.
\newblock \bibinfo{journal}{\emph{arXiv preprint arXiv:2308.10835}} (\bibinfo{year}{2023}).
\newblock


\bibitem[Wang et~al\mbox{.}(2025c)]%
        {wang2025graph}
\bibfield{author}{\bibinfo{person}{Zehong Wang}, \bibinfo{person}{Zheyuan Liu}, \bibinfo{person}{Tianyi Ma}, \bibinfo{person}{Jiazheng Li}, \bibinfo{person}{Zheyuan Zhang}, \bibinfo{person}{Xingbo Fu}, \bibinfo{person}{Yiyang Li}, \bibinfo{person}{Zhengqing Yuan}, \bibinfo{person}{Wei Song}, \bibinfo{person}{Yijun Ma}, {et~al\mbox{.}}} \bibinfo{year}{2025}\natexlab{c}.
\newblock \showarticletitle{Graph Foundation Models: A Comprehensive Survey}.
\newblock \bibinfo{journal}{\emph{arXiv preprint arXiv:2505.15116}} (\bibinfo{year}{2025}).
\newblock


\bibitem[Wei et~al\mbox{.}(2024)]%
        {wei2024llmrec}
\bibfield{author}{\bibinfo{person}{Wei Wei}, \bibinfo{person}{Xubin Ren}, \bibinfo{person}{Jiabin Tang}, \bibinfo{person}{Qinyong Wang}, \bibinfo{person}{Lixin Su}, \bibinfo{person}{Suqi Cheng}, \bibinfo{person}{Junfeng Wang}, \bibinfo{person}{Dawei Yin}, {and} \bibinfo{person}{Chao Huang}.} \bibinfo{year}{2024}\natexlab{}.
\newblock \showarticletitle{Llmrec: Large language models with graph augmentation for recommendation}. In \bibinfo{booktitle}{\emph{Proceedings of the 17th ACM International Conference on Web Search and Data Mining}}. \bibinfo{pages}{806--815}.
\newblock


\bibitem[Wei et~al\mbox{.}(2025)]%
        {wei2025mscrs}
\bibfield{author}{\bibinfo{person}{Yibiao Wei}, \bibinfo{person}{Jie Zou}, \bibinfo{person}{Weikang Guo}, \bibinfo{person}{Guoqing Wang}, \bibinfo{person}{Xing Xu}, {and} \bibinfo{person}{Yang Yang}.} \bibinfo{year}{2025}\natexlab{}.
\newblock \showarticletitle{MSCRS: Multi-modal semantic graph prompt learning framework for conversational recommender systems}. In \bibinfo{booktitle}{\emph{Proceedings of the 48th International ACM SIGIR Conference on Research and Development in Information Retrieval}}. \bibinfo{pages}{42--52}.
\newblock


\bibitem[Wu et~al\mbox{.}(2024a)]%
        {wu2024exploring}
\bibfield{author}{\bibinfo{person}{Likang Wu}, \bibinfo{person}{Zhaopeng Qiu}, \bibinfo{person}{Zhi Zheng}, \bibinfo{person}{Hengshu Zhu}, {and} \bibinfo{person}{Enhong Chen}.} \bibinfo{year}{2024}\natexlab{a}.
\newblock \showarticletitle{Exploring large language model for graph data understanding in online job recommendations}. In \bibinfo{booktitle}{\emph{Proceedings of the AAAI Conference on Artificial Intelligence}}, Vol.~\bibinfo{volume}{38}. \bibinfo{pages}{9178--9186}.
\newblock


\bibitem[Wu et~al\mbox{.}(2024b)]%
        {wu2024survey}
\bibfield{author}{\bibinfo{person}{Likang Wu}, \bibinfo{person}{Zhi Zheng}, \bibinfo{person}{Zhaopeng Qiu}, \bibinfo{person}{Hao Wang}, \bibinfo{person}{Hongchao Gu}, \bibinfo{person}{Tingjia Shen}, \bibinfo{person}{Chuan Qin}, \bibinfo{person}{Chen Zhu}, \bibinfo{person}{Hengshu Zhu}, \bibinfo{person}{Qi Liu}, {et~al\mbox{.}}} \bibinfo{year}{2024}\natexlab{b}.
\newblock \showarticletitle{A survey on large language models for recommendation}.
\newblock \bibinfo{journal}{\emph{World Wide Web}} \bibinfo{volume}{27}, \bibinfo{number}{5} (\bibinfo{year}{2024}), \bibinfo{pages}{60}.
\newblock


\bibitem[Wu et~al\mbox{.}(2022)]%
        {wu2022graph}
\bibfield{author}{\bibinfo{person}{Shiwen Wu}, \bibinfo{person}{Fei Sun}, \bibinfo{person}{Wentao Zhang}, \bibinfo{person}{Xu Xie}, {and} \bibinfo{person}{Bin Cui}.} \bibinfo{year}{2022}\natexlab{}.
\newblock \showarticletitle{Graph neural networks in recommender systems: a survey}.
\newblock \bibinfo{journal}{\emph{Comput. Surveys}} \bibinfo{volume}{55}, \bibinfo{number}{5} (\bibinfo{year}{2022}), \bibinfo{pages}{1--37}.
\newblock


\bibitem[Xu et~al\mbox{.}(2025)]%
        {Xu_Wei_Yang_Wu_2025}
\bibfield{author}{\bibinfo{person}{Mingtao Xu}, \bibinfo{person}{Wei Wei}, \bibinfo{person}{Peixuan Yang}, {and} \bibinfo{person}{Hulong Wu}.} \bibinfo{year}{2025}\natexlab{}.
\newblock \showarticletitle{Semantic Enhanced Heterogeneous Hypergraph Network for Collaborative Filtering}.
\newblock \bibinfo{journal}{\emph{Proceedings of the AAAI Conference on Artificial Intelligence}} \bibinfo{volume}{39}, \bibinfo{number}{12} (\bibinfo{date}{Apr.} \bibinfo{year}{2025}), \bibinfo{pages}{12936--12944}.
\newblock


\bibitem[Xu and Zhang(2025)]%
        {xu2025llm}
\bibfield{author}{\bibinfo{person}{Zekun Xu} {and} \bibinfo{person}{Yudi Zhang}.} \bibinfo{year}{2025}\natexlab{}.
\newblock \showarticletitle{LLM-Enhanced Reranking for Complementary Product Recommendation}.
\newblock \bibinfo{journal}{\emph{arXiv preprint arXiv:2507.16237}} (\bibinfo{year}{2025}).
\newblock


\bibitem[Yang et~al\mbox{.}(2023)]%
        {yang2023palr}
\bibfield{author}{\bibinfo{person}{Fan Yang}, \bibinfo{person}{Zheng Chen}, \bibinfo{person}{Ziyan Jiang}, \bibinfo{person}{Eunah Cho}, \bibinfo{person}{Xiaojiang Huang}, {and} \bibinfo{person}{Yanbin Lu}.} \bibinfo{year}{2023}\natexlab{}.
\newblock \showarticletitle{Palr: Personalization aware llms for recommendation}.
\newblock \bibinfo{journal}{\emph{arXiv preprint arXiv:2305.07622}} (\bibinfo{year}{2023}).
\newblock


\bibitem[Yang et~al\mbox{.}(2025a)]%
        {yang2025training}
\bibfield{author}{\bibinfo{person}{Mingdai Yang}, \bibinfo{person}{Zhiwei Liu}, \bibinfo{person}{Liangwei Yang}, \bibinfo{person}{Xiaolong Liu}, \bibinfo{person}{Chen Wang}, \bibinfo{person}{Hao Peng}, {and} \bibinfo{person}{Philip~S Yu}.} \bibinfo{year}{2025}\natexlab{a}.
\newblock \showarticletitle{Training Large Recommendation Models via Graph-Language Tokens Alignment}. In \bibinfo{booktitle}{\emph{Companion Proceedings of the ACM on Web Conference 2025}}. \bibinfo{pages}{1470--1474}.
\newblock


\bibitem[Yang et~al\mbox{.}(2024a)]%
        {yang2024sequential}
\bibfield{author}{\bibinfo{person}{Shenghao Yang}, \bibinfo{person}{Weizhi Ma}, \bibinfo{person}{Peijie Sun}, \bibinfo{person}{Qingyao Ai}, \bibinfo{person}{Yiqun Liu}, \bibinfo{person}{Mingchen Cai}, {and} \bibinfo{person}{Min Zhang}.} \bibinfo{year}{2024}\natexlab{a}.
\newblock \showarticletitle{Sequential recommendation with latent relations based on large language model}. In \bibinfo{booktitle}{\emph{Proceedings of the 47th International ACM SIGIR Conference on Research and Development in Information Retrieval}}. \bibinfo{pages}{335--344}.
\newblock


\bibitem[Yang et~al\mbox{.}(2024b)]%
        {yang2024common}
\bibfield{author}{\bibinfo{person}{Shenghao Yang}, \bibinfo{person}{Weizhi Ma}, \bibinfo{person}{Peijie Sun}, \bibinfo{person}{Min Zhang}, \bibinfo{person}{Qingyao Ai}, \bibinfo{person}{Yiqun Liu}, {and} \bibinfo{person}{Mingchen Cai}.} \bibinfo{year}{2024}\natexlab{b}.
\newblock \showarticletitle{Common sense enhanced knowledge-based recommendation with large language model}. In \bibinfo{booktitle}{\emph{International Conference on Database Systems for Advanced Applications}}. Springer, \bibinfo{pages}{381--390}.
\newblock


\bibitem[Yang et~al\mbox{.}(2025b)]%
        {yang2025cold}
\bibfield{author}{\bibinfo{person}{Wooseong Yang}, \bibinfo{person}{Weizhi Zhang}, \bibinfo{person}{Yuqing Liu}, \bibinfo{person}{Yuwei Han}, \bibinfo{person}{Yu Wang}, \bibinfo{person}{Junhyun Lee}, {and} \bibinfo{person}{Philip~S Yu}.} \bibinfo{year}{2025}\natexlab{b}.
\newblock \showarticletitle{Cold-Start Recommendation with Knowledge-Guided Retrieval-Augmented Generation}.
\newblock \bibinfo{journal}{\emph{arXiv preprint arXiv:2505.20773}} (\bibinfo{year}{2025}).
\newblock


\bibitem[Yang and Rahmani(2025)]%
        {yang2025personalized}
\bibfield{author}{\bibinfo{person}{Zhongqi Yang} {and} \bibinfo{person}{Amir Rahmani}.} \bibinfo{year}{2025}\natexlab{}.
\newblock \showarticletitle{Personalized Causal Graph Reasoning for LLMs: A Case Study on Dietary Recommendations}.
\newblock \bibinfo{journal}{\emph{arXiv preprint arXiv:2503.00134}} (\bibinfo{year}{2025}).
\newblock


\bibitem[Yong et~al\mbox{.}(2025)]%
        {yong2025enhancing}
\bibfield{author}{\bibinfo{person}{Qian Yong}, \bibinfo{person}{Yanhui Li}, \bibinfo{person}{Jialiang Shi}, \bibinfo{person}{Yaguang Dou}, {and} \bibinfo{person}{Tian Qi}.} \bibinfo{year}{2025}\natexlab{}.
\newblock \showarticletitle{Enhancing Serendipity Recommendation System by Constructing Dynamic User Knowledge Graphs with Large Language Models}.
\newblock \bibinfo{journal}{\emph{arXiv preprint arXiv:2508.04032}} (\bibinfo{year}{2025}).
\newblock


\bibitem[Yu et~al\mbox{.}({[n.\,d.]})]%
        {yuyou}
\bibfield{author}{\bibinfo{person}{Qing Yu}, \bibinfo{person}{Xiaobei Wang}, \bibinfo{person}{Shuchang Liu}, \bibinfo{person}{Xiaoyu Yang}, \bibinfo{person}{Xueliang Wang}, \bibinfo{person}{Chang Meng}, \bibinfo{person}{Shanshan Wu}, \bibinfo{person}{Bin Wen}, \bibinfo{person}{Huihui Xiao}, \bibinfo{person}{Xiang Li}, {et~al\mbox{.}}} \bibinfo{year}{[n.\,d.]}\natexlab{}.
\newblock \showarticletitle{Who You Are Matters: Bridging Interests and Social Roles via LLM-Enhanced Logic Recommendation}. In \bibinfo{booktitle}{\emph{The Thirty-ninth Annual Conference on Neural Information Processing Systems}}.
\newblock


\bibitem[Zhai et~al\mbox{.}(2024)]%
        {zhai2024actions}
\bibfield{author}{\bibinfo{person}{Jiaqi Zhai}, \bibinfo{person}{Lucy Liao}, \bibinfo{person}{Xing Liu}, \bibinfo{person}{Yueming Wang}, \bibinfo{person}{Rui Li}, \bibinfo{person}{Xuan Cao}, \bibinfo{person}{Leon Gao}, \bibinfo{person}{Zhaojie Gong}, \bibinfo{person}{Fangda Gu}, \bibinfo{person}{Jiayuan He}, \bibinfo{person}{Yinghai Lu}, {and} \bibinfo{person}{Yu Shi}.} \bibinfo{year}{2024}\natexlab{}.
\newblock \showarticletitle{Actions Speak Louder than Words: Trillion-Parameter Sequential Transducers for Generative Recommendations}. In \bibinfo{booktitle}{\emph{Forty-first International Conference on Machine Learning}}.
\newblock


\bibitem[Zhang et~al\mbox{.}(2023)]%
        {zhang2023robust}
\bibfield{author}{\bibinfo{person}{Kaike Zhang}, \bibinfo{person}{Qi Cao}, \bibinfo{person}{Fei Sun}, \bibinfo{person}{Yunfan Wu}, \bibinfo{person}{Shuchang Tao}, \bibinfo{person}{Huawei Shen}, {and} \bibinfo{person}{Xueqi Cheng}.} \bibinfo{year}{2023}\natexlab{}.
\newblock \showarticletitle{Robust recommender system: a survey and future directions}.
\newblock \bibinfo{journal}{\emph{arXiv preprint arXiv:2309.02057}} (\bibinfo{year}{2023}).
\newblock


\bibitem[Zhang et~al\mbox{.}(2025d)]%
        {zhang2025kedrec}
\bibfield{author}{\bibinfo{person}{Kai Zhang}, \bibinfo{person}{Rui Zhu}, \bibinfo{person}{Shutian Ma}, \bibinfo{person}{Jingwei Xiong}, \bibinfo{person}{Yejin Kim}, \bibinfo{person}{Fabricio Murai}, {and} \bibinfo{person}{Xiaozhong Liu}.} \bibinfo{year}{2025}\natexlab{d}.
\newblock \showarticletitle{Kedrec-lm: A knowledge-distilled explainable drug recommendation large language model}.
\newblock \bibinfo{journal}{\emph{arXiv preprint arXiv:2502.20350}} (\bibinfo{year}{2025}).
\newblock


\bibitem[Zhang et~al\mbox{.}(2025b)]%
        {zhang2025integrating}
\bibfield{author}{\bibinfo{person}{Shuo Zhang}, \bibinfo{person}{Xiao Li}, \bibinfo{person}{Jiayi Wu}, \bibinfo{person}{Fan Yang}, \bibinfo{person}{Xiang Li}, {and} \bibinfo{person}{Ming Gao}.} \bibinfo{year}{2025}\natexlab{b}.
\newblock \showarticletitle{Integrating LLM-Derived Multi-Semantic Intent into Graph Model for Session-based Recommendation}.
\newblock \bibinfo{journal}{\emph{arXiv preprint arXiv:2507.20147}} (\bibinfo{year}{2025}).
\newblock


\bibitem[Zhang et~al\mbox{.}(2025c)]%
        {zhang-etal-2025-llminit}
\bibfield{author}{\bibinfo{person}{Weizhi Zhang}, \bibinfo{person}{Liangwei Yang}, \bibinfo{person}{Wooseong Yang}, \bibinfo{person}{Henry~Peng Zou}, \bibinfo{person}{Yuqing Liu}, \bibinfo{person}{Ke Xu}, \bibinfo{person}{Sourav Medya}, {and} \bibinfo{person}{Philip~S. Yu}.} \bibinfo{year}{2025}\natexlab{c}.
\newblock \showarticletitle{{LLMI}nit: A Free Lunch from Large Language Models for Selective Initialization of Recommendation}. In \bibinfo{booktitle}{\emph{Proceedings of the 2025 Conference on Empirical Methods in Natural Language Processing: Industry Track}}, \bibfield{editor}{\bibinfo{person}{Saloni Potdar}, \bibinfo{person}{Lina Rojas-Barahona}, {and} \bibinfo{person}{Sebastien Montella}} (Eds.). \bibinfo{publisher}{Association for Computational Linguistics}, \bibinfo{address}{Suzhou (China)}, \bibinfo{pages}{2016--2024}.
\newblock
\showISBNx{979-8-89176-333-3}


\bibitem[Zhang et~al\mbox{.}(2025a)]%
        {zhang-etal-2025-bi}
\bibfield{author}{\bibinfo{person}{Xinyu Zhang}, \bibinfo{person}{Linmei Hu}, \bibinfo{person}{Luhao Zhang}, \bibinfo{person}{Wentao Cheng}, \bibinfo{person}{Yashen Wang}, \bibinfo{person}{Ge Shi}, \bibinfo{person}{Chong Feng}, {and} \bibinfo{person}{Liqiang Nie}.} \bibinfo{year}{2025}\natexlab{a}.
\newblock \showarticletitle{Bi-Tuning with Collaborative Information for Controllable {LLM}-based Sequential Recommendation}. In \bibinfo{booktitle}{\emph{Proceedings of the 63rd Annual Meeting of the Association for Computational Linguistics (Volume 1: Long Papers)}}, \bibfield{editor}{\bibinfo{person}{Wanxiang Che}, \bibinfo{person}{Joyce Nabende}, \bibinfo{person}{Ekaterina Shutova}, {and} \bibinfo{person}{Mohammad~Taher Pilehvar}} (Eds.). \bibinfo{publisher}{Association for Computational Linguistics}, \bibinfo{address}{Vienna, Austria}, \bibinfo{pages}{19340--19351}.
\newblock
\showISBNx{979-8-89176-251-0}


\bibitem[Zhang et~al\mbox{.}(2024)]%
        {zhang2024finerec}
\bibfield{author}{\bibinfo{person}{Xiaokun Zhang}, \bibinfo{person}{Bo Xu}, \bibinfo{person}{Youlin Wu}, \bibinfo{person}{Yuan Zhong}, \bibinfo{person}{Hongfei Lin}, {and} \bibinfo{person}{Fenglong Ma}.} \bibinfo{year}{2024}\natexlab{}.
\newblock \showarticletitle{FineRec: Exploring Fine-grained Sequential Recommendation}. In \bibinfo{booktitle}{\emph{Proceedings of the 47th International ACM SIGIR Conference on Research and Development in Information Retrieval}}. \bibinfo{pages}{1599--1608}.
\newblock


\bibitem[Zhao et~al\mbox{.}(2025)]%
        {zhao2025reason}
\bibfield{author}{\bibinfo{person}{Keyu Zhao}, \bibinfo{person}{Fengli Xu}, {and} \bibinfo{person}{Yong Li}.} \bibinfo{year}{2025}\natexlab{}.
\newblock \showarticletitle{Reason-to-Recommend: Using Interaction-of-Thought Reasoning to Enhance LLM Recommendation}.
\newblock \bibinfo{journal}{\emph{arXiv preprint arXiv:2506.05069}} (\bibinfo{year}{2025}).
\newblock


\bibitem[Zhao et~al\mbox{.}(2024b)]%
        {zhao2024breaking}
\bibfield{author}{\bibinfo{person}{Qian Zhao}, \bibinfo{person}{Hao Qian}, \bibinfo{person}{Ziqi Liu}, \bibinfo{person}{Gong-Duo Zhang}, {and} \bibinfo{person}{Lihong Gu}.} \bibinfo{year}{2024}\natexlab{b}.
\newblock \showarticletitle{Breaking the barrier: utilizing large language models for industrial recommendation systems through an inferential knowledge graph}. In \bibinfo{booktitle}{\emph{Proceedings of the 33rd ACM International Conference on Information and Knowledge Management}}. \bibinfo{pages}{5086--5093}.
\newblock


\bibitem[Zhao et~al\mbox{.}(2024a)]%
        {zhao2024dynllm}
\bibfield{author}{\bibinfo{person}{Ziwei Zhao}, \bibinfo{person}{Fake Lin}, \bibinfo{person}{Xi Zhu}, \bibinfo{person}{Zhi Zheng}, \bibinfo{person}{Tong Xu}, \bibinfo{person}{Shitian Shen}, \bibinfo{person}{Xueying Li}, \bibinfo{person}{Zikai Yin}, {and} \bibinfo{person}{Enhong Chen}.} \bibinfo{year}{2024}\natexlab{a}.
\newblock \showarticletitle{DynLLM: When Large Language Models Meet Dynamic Graph Recommendation}.
\newblock \bibinfo{journal}{\emph{arXiv preprint arXiv:2405.07580}} (\bibinfo{year}{2024}).
\newblock


\bibitem[Zheng et~al\mbox{.}({[n.\,d.]})]%
        {zhengtuning}
\bibfield{author}{\bibinfo{person}{Wenqing Zheng}, \bibinfo{person}{Noah Fatsi}, \bibinfo{person}{Daniel Barcklow}, \bibinfo{person}{Dmitri Kalaev}, \bibinfo{person}{Steven Yao}, \bibinfo{person}{Owen Reinert}, \bibinfo{person}{C~Bayan Bruss}, {and} \bibinfo{person}{Daniele Rosa}.} \bibinfo{year}{[n.\,d.]}\natexlab{}.
\newblock \showarticletitle{Tuning-Free LLM Can Build A Strong Recommender Under Sparse Connectivity And Knowledge Gap Via Extracting Intent}. In \bibinfo{booktitle}{\emph{The Fourth Learning on Graphs Conference}}.
\newblock


\bibitem[Zhou et~al\mbox{.}(2025)]%
        {zhou-etal-2025-efficiency}
\bibfield{author}{\bibinfo{person}{Huixue Zhou}, \bibinfo{person}{Hengrui Gu}, \bibinfo{person}{Zaifu Zhan}, \bibinfo{person}{Xi Liu}, \bibinfo{person}{Kaixiong Zhou}, \bibinfo{person}{Yongkang Xiao}, \bibinfo{person}{Mingfu Liang}, \bibinfo{person}{Srinivas~Prasad Govindan}, \bibinfo{person}{Piyush Chawla}, \bibinfo{person}{Jiyan Yang}, \bibinfo{person}{Xiangfei Meng}, \bibinfo{person}{Huayu Li}, \bibinfo{person}{Buyun Zhang}, \bibinfo{person}{Liang Luo}, \bibinfo{person}{Wen-Yen Chen}, \bibinfo{person}{Yiping Han}, \bibinfo{person}{Bo Long}, \bibinfo{person}{Rui Zhang}, {and} \bibinfo{person}{Tianlong Chen}.} \bibinfo{year}{2025}\natexlab{}.
\newblock \showarticletitle{The Efficiency vs. Accuracy Trade-off: Optimizing {RAG}-Enhanced {LLM} Recommender Systems Using Multi-Head Early Exit}. In \bibinfo{booktitle}{\emph{Proceedings of the 63rd Annual Meeting of the Association for Computational Linguistics (Volume 1: Long Papers)}}, \bibfield{editor}{\bibinfo{person}{Wanxiang Che}, \bibinfo{person}{Joyce Nabende}, \bibinfo{person}{Ekaterina Shutova}, {and} \bibinfo{person}{Mohammad~Taher Pilehvar}} (Eds.). \bibinfo{publisher}{Association for Computational Linguistics}, \bibinfo{address}{Vienna, Austria}, \bibinfo{pages}{26443--26458}.
\newblock
\showISBNx{979-8-89176-251-0}


\end{thebibliography}



\end{document}